\documentclass[10pt]{article}
\usepackage[a4paper,margin=1in]{geometry}

\usepackage{jheppub}

\usepackage[T1]{fontenc}

\usepackage{graphicx}
\usepackage{xcolor}
\usepackage{amssymb}
\usepackage{amsmath}
\usepackage{dsfont}
\usepackage{setspace}
\usepackage{slashed}
\usepackage{pifont}
\usepackage[capitalize]{cleveref}
\usepackage{multirow}
\usepackage{tabularx}
\newcolumntype{C}[1]{>{\hsize=#1\hsize\centering\arraybackslash}X}%

\newcolumntype{Z}{r<{\hspace{3mm}}}

\renewcommand{\arraystretch}{1.4}

\def\cA{\mathcal{A}}
\def\cM{\mathcal{M}}
\def\cF{\mathcal{F}}
\def\cR{\mathcal{R}}
\def\cH{\mathcal{H}}
\def\cO{\mathcal{O}}
\def\chQ{\mathcal{Q}}
\def\eps{\epsilon}
\DeclareMathOperator{\tr}{\rm tr}
\def\nn{\nonumber \\ }
\def\trm{\tr_-}
\def\trp{\tr_+}

\def\qqgga{0\to \bar{q}qgg\gamma}
\def\qqQQa{0\to q\bar{q}Q\bar{Q}\gamma}
\def\qqqqa{0\to q\bar{q}q\bar{q}\gamma}

\def\qqggaX{\bar{q}qgg\gamma}
\def\qqQQaX{q\bar{q}Q\bar{Q}\gamma}

\def\la{\langle}
\def\ra{\rangle}
\def\spA#1#2{\la#1#2\ra}
\def\spB#1#2{[#1#2]}

\def\colhelsum{\underset{\mathrm{colour}}{\overline{\sum}}\,\underset{\mathrm{helicity}}{\overline{\sum}}}

\definecolor{mygreen}{rgb}{0,0.7,0}

\title{Isolated photon production in association with a jet pair through next-to-next-to-leading order in QCD}

\author[a]{Simon Badger,}
\author[b]{Micha\l{} Czakon,}
\author[c]{Heribertus Bayu Hartanto,}
\author[a]{Ryan Moodie,}
\author[d]{Tiziano Peraro,}
\author[c]{Rene Poncelet,}
\author[a]{Simone Zoia}

\affiliation[a]{Dipartimento di Fisica and Arnold-Regge Center, Università di Torino, and INFN, Sezione di Torino, Via P. Giuria 1, I-10125 Torino, Italy}
\affiliation[b]{Institut f\"ur Theoretische Teilchenphysik und Kosmologie, RWTH Aachen University,\\ D-52056 Aachen, Germany}
\affiliation[c]{Cavendish Laboratory, University of Cambridge, Cambridge CB3 0HE, United Kingdom}
\affiliation[d]{Dipartimento di Fisica e Astronomia, Università di Bologna e INFN, Sezione di Bologna, via Irnerio 46, I-40126 Bologna, Italy}

\emailAdd{simondavid.badger@unito.it}
\emailAdd{mczakon@physik.rwth-aachen.de}
\emailAdd{hbhartanto@hep.phy.cam.ac.uk}
\emailAdd{ryaniain.moodie@unito.it}
\emailAdd{tiziano.peraro@unibo.it}
\emailAdd{poncelet@hep.phy.cam.ac.uk}
\emailAdd{simone.zoia@unito.it}

\preprint{Cavendish-HEP-23/02, P3H-23-022, TTK-23-09}

\abstract{In this work, we provide a comprehensive set of differential cross-section distributions for photon + di-jet production in proton-proton collisions with next-to-next-to-leading order precision in massless QCD. The event selection corresponds to recent measurements by the ATLAS collaboration. We observe an improved description of data in comparison to lower-order calculations in the case of observables that are expected to be well described by perturbation theory. The results also show better agreement with data than parton-shower-matched and multi-jet-merged predictions generated for the ATLAS analysis using the \textsc{Sherpa} Monte Carlo. A particular highlight of our study is the use of exact five-point two-loop virtual amplitudes. This is the first calculation of a complete two-to-three hadron-collider process at next-to-next-to-leading order in QCD that does not rely on the leading-colour approximation at two loops. We demonstrate, nevertheless, that the sub-leading-colour effects present in the infrared- and ultraviolet-finite double-virtual contributions are negligible in view of the remaining scale uncertainties.}

\begin{document}
\maketitle
\flushbottom

\section{Introduction}
\label{sec:introduction}

The production of photons in high-energy collisions has a rich phenomenology due to prompt production in the partonic process as well as secondary emission after the hadronization phase. In fact, the most significant fraction of photons produced at colliders such as the Large Hadron Collider (LHC) originate from hadron decays. The smaller fraction of prompt (or primary) photons, however, is more interesting to us since those photons allow for the application and testing of perturbative Quantum Chromodynamics (QCD). Prompt photons can be produced through two mechanisms: hard emission and fragmentation. Photons produced by the first mechanism are sensitive to the QCD dynamics of the interaction and are well separated from the other participating partons by definition. In contrast, photons produced through fragmentation of the partonic final state are found in proximity to hadrons, since they are generated by collinear emissions from partons. The separation is not unique, since it requires to define the exact meaning of collinear emissions.

In order to separate the primary from the secondary photons, isolation criteria are implemented experimentally, restricting the hadronic activity around a photon candidate by requiring the (transverse) hadronic energy, $E_{\perp}$, in a cone of size $R$ to be below (a possibly dynamical) threshold. A non-zero threshold, as practical realities require, implies a non-vanishing fragmentation component and, therefore, requires a corresponding treatment of these contributions in theory predictions. The fragmentation process can be treated in collinear/mass factorization \cite{Koller:1978kq, Laermann:1982jr}. It has been implemented in public codes up to NLO, for example in \textsc{JetPhox} \cite{Catani:2002ny}. Recently, first predictions at NNLO QCD, including photon fragmentation, have become available \cite{Gehrmann:2022cih, Chen:2022gpk}. Within the fragmentation formalism, collinear divergencies are absorbed into a fragmentation function that contains perturbative and non-perturbative contributions. The latter must either be modelled or measured \cite{Glover:1993xc, ALEPH:1995zdi, Bourhis:1997yu, OPAL:1997lep, Belghobsi:2009hx, Kaufmann:2017lsd}. Alternatively, the so-called smooth-cone isolation, or Fixione isolation, criterion \cite{Frixione:1998jh} can be implemented to entirely remove the fragmentation component from the theory predictions by vetoing radiation collinear to the photon. This type of isolation has been used in many QCD studies \cite{Catani:2002ny, Gehrmann:2013aga, Chen:2019zmr, Gehrmann:2020oec, Chen:2022gpk}. Frixione isolation is handy from the theory point of view but cannot be implemented in experiment due to the limited resolution of detectors. A compromise is achieved by the usage of hybrid isolation \cite{Siegert:2016bre, Chen:2022gpk}, which combines a smooth-cone and a hard-cone prescription. While this is closer to the experimental setup, the removed fragmentation component can still lead to corrections at the percent level \cite{Chen:2022gpk}. It should be mentioned that isolation criteria lead to logarithmically enhanced contributions for small isolation regions. This may induce sizeable resummation effects \cite{Catani:2013oma, Balsiger:2018ezi, Becher:2022rhu}.

The production of isolated photons has been studied experimentally since early days of collider \cite{Athens-Athens-Brookhaven-CERN:1979oxx, Anassontzis:1982gm, CMOR:1989qzc, UA1:1988zam, UA2:1991wce} and fixed-target experiments \cite{WA70:1987vvj, FermilabE706:2004emk}. At the Tevatron, for example, cross sections for photons and photons accompanied by jets have been extensively measured as demonstrated by Refs.~\cite{D0:2005ofv, D0:2008chx, D0:2013lra, CDF:2017cuc}. More recently, isolated photon production has become one of the standard-candle measurements at the LHC \cite{ATLAS:2017nah, ATLAS:2017xqp, ATLAS:2019buk, CMS:2018qao, CMS:2019jlq, ALICE:2019rtd, ATLAS:2019iaa}. Inclusive production of a photon and the production of a photon with a single resolved jet are particularly sensitive to the gluon content of the proton since for a non-vanishing transverse momentum of the photon the dominant partonic-process is $qg \to \gamma q$. Furthermore, these processes can also be used for the tuning of Monte Carlo generators as well as for jet-energy calibration \cite{Belghobsi:2009hx}.

A photon associated with a pair of jets has some unique features, making it particularly interesting for precision QCD phenomenology \cite{Keller:1991xc}. Due to the $2 \to 3$ kinematics, it can be used to study angular correlations between the photon and jets similarly to multi-jet production processes. However, in contrast to the three-jet case, the photon is identified, and the phase space can be subdivided into three regions:
\begin{itemize}
    \item $E_{\perp}(\gamma) > p_T(j_1)$: dominated by diagrams of $pp \to \gamma q$-type with an additional radiated gluon. This region has enhanced contributions from photons originating from the hard interaction.
    \item $p_T(j_1) > E_{\perp}(\gamma) > p_T(j_2)$: dominated by ``Bremsstrahlung'' di-jet production $pp \to jj$ where one final state quark radiates a high energy photon. This region is sensitive to the high $z$ region of the photon fragmentation process, with $z$ the collinear-momentum fraction of the photon.
    \item $p_T(j_2) > E_{\perp}(\gamma)$: the ``Bremsstrahlung'' configuration with further soft-photon enhancement. This region is sensitive to the lower $z$ region of the photon fragmentation.
\end{itemize}
Besides providing a test of perturbative QCD and the fragmentation formalism, the process we wish to study allows us to assess and tune Monte-Carlo event generators. In this respect it is similar to processes with a lower number of jets. It is, however, more challenging as far as modelling of jet activity is concerned. As a last application, we also mention that photon + di-jet production allows for the calibration of data-driven background estimates \cite{Lindert:2017olm}.

Higher-order corrections to the production cross-section of an isolated photon with two jets have been computed through NLO QCD  and matched to parton-shower simulations \cite{Gleisberg:2008ta, Sherpa:2019gpd}. NNLO QCD cross sections for $2\to3$ processes \cite{Chawdhry:2019bji,  Kallweit:2020gcp,  Chawdhry:2021hkp, Czakon:2021mjy, Badger:2021ohm, Hartanto:2022qhh, Hartanto:2022ypo, Buonocore:2022pqq, Alvarez:2023fhi} are the current state-of-the-art in perturbative QCD. The main bottleneck is the computation of five-point two-loop amplitudes. In the case of massless final states, all relevant partonic processes for three-jet \cite{Abreu:2019odu, Abreu:2021oya}, three-photon \cite{Chawdhry:2019bji, Abreu:2020cwb, Chawdhry:2020for}, and di-photon + jet \cite{Agarwal:2021grm, Chawdhry:2021mkw} production have been computed in leading-colour approximation. These computations represent the appearing Feynman integrals in terms of {\it pentagon functions} \cite{Chicherin:2018old, Chicherin:2020oor}, which allow for a compact analytical form and a stable numerical evaluation. The function space for the case of a single external mass has also been worked out, and the first leading-colour amplitudes involving an external massive vector boson \cite{Hartanto:2019uvl, Badger:2021imn, Badger:2021nhg, Badger:2021ega, Badger:2022ncb} have been computed. The pentagon functions are also available for the non-planar integrals appearing beyond the leading-colour approximation. Still, only the di-photon + jet amplitude has been computed at full colour \cite{Agarwal:2021vdh}. In this article, we evaluate all partonic amplitudes needed for the photon + di-jet cross sections at full colour.
Equipped with the amplitudes, we perform a first phenomenological study of the process matching the configuration of recent measurements by ATLAS \cite{ATLAS:2019iaa}.

The article is organised as follows. In \cref{sec:amplitude}, we describe the methods used for the evaluation of the two-loop contributions. After introducing the kinematic variables, we provide the helicity- and colour-space decomposition of the amplitudes. The computation of the latter by means of integration-by-parts (IBP) identities improved by syzygy equations is described, as well as the transformations of the occurring special functions necessary to obtain the amplitudes for all configurations. The section is closed with comments on the numerical evaluation including methods to improve numerical stability. In \cref{sec:phenomenology}, we provide differential cross-section distributions for a number of observables for a setup used in a recent measurement by the ATLAS collaboration. We comment on the convergence of the perturbative series and discuss the possible effects of omitted electroweak and fragmentation effects. We close the main text with conclusions and outlook on future work. In \cref{app:benchmark}, we provide numerical values of the amplitudes for a benchmark phase-space point which allows for an easy verification of our results by independent parties. In \cref{app:mu}, we reproduce explicit formulae for the scale dependence of the amplitudes, which is necessary since our results are derived for one particular value of the renormalisation scale. Finally, in \cref{app:parity}, we describe the parity transformation and external-momentum permutations necessary to derive the complete amplitudes from partial results for specific helicity configurations.

\section{Two-loop amplitudes}\label{sec:amplitude}
The computation of $pp \to \gamma jj$ production at NNLO QCD accuracy requires one- and two-loop QCD corrections to the following partonic processes:
\begin{itemize}
    \item 2-quark-2-gluon channel ($\qqgga$),
    \item 4-quark channel ($\qqQQa$ with either $Q = q$ or $Q \neq q$).
\end{itemize}
In addition to these, the $0\to gggg\gamma$ partonic channel also enters starting at NNLO QCD, since it is a loop-induced process where the leading order contribution is made of one-loop amplitudes. In this work, we compute analytically the one- and two-loop QCD corrections to $\qqgga$ and $\qqQQa$.\footnote{We compute the one-loop amplitudes including the higher-order terms in the dimensional regulator parameter $\eps$.} 
The one-loop amplitude for $0\to gggg\gamma$, as well the amplitudes contributing to the real-virtual corrections 
to the NNLO QCD cross section (i.e.\ the one-loop amplitudes for $pp\to\gamma jjj$) are obtained from \textsc{OpenLoops}~\cite{Buccioni:2019sur}. 

\subsection{Kinematics}
We take the external momenta $p_i$ in the five-particle processes $\qqgga$ and $\qqQQa$ to be outgoing. They satisfy momentum conservation,
\begin{equation}
\sum_{i=1}^{5} p_i = 0 \,,
\end{equation}
as well as the on-shell conditions $p_i^2 = 0$ for $i=1,\ldots,5$.
The five-particle phase space is described by five independent scalar invariants,
\begin{equation}
\vec{s} = \left\lbrace s_{12},s_{23},s_{34},s_{45},s_{15} \right\rbrace \,,
\label{eq:sijs}
\end{equation}
with $s_{ij} = (p_i + p_j)^2$, and a pseudo-scalar quantity,
\begin{equation} \label{eq:tr5}
\tr_5 = 4 \mathrm{i} \eps_{\mu\nu\rho\sigma} p_1^\mu p_2^\nu p_3^\rho p_4^\sigma = 
\spB{1}{2}\spA{2}{3}\spB{3}{4}\spA{4}{1} - \spA{1}{2}\spB{2}{3}\spA{3}{4}\spB{4}{1} \,.
\end{equation}
The square of the latter is a scalar quantity, and can thus be expressed in terms of $\vec{s}$. We have $\tr_5^2 = \Delta_5$, where $\Delta_5$ is the Gram determinant of the external momenta,
\begin{align}
\Delta_5 = \text{det}\left( s_{ij} \right)_{i,j=1,\ldots,4} \,,
\end{align}
which is a degree-4 polynomial in the $\vec{s}$. The parameterisation of the five-particle kinematics therefore contains a square root: $\sqrt{\Delta_5}$. This is important in view of the computation with finite field arithmetic discussed in \cref{sec:amplitude_computation}, which requires a rational parameterisation of the kinematics.

We work in dimensional regularisation, with $d=4-2 \eps$ spacetime dimensions, and keep the external momenta four-dimensional. This allows us to use the spinor-helicity formalism, as shown in 
\cref{eq:tr5}. We set the renormalisation scale $\mu_R$ to $1$ throughout the computation of the amplitudes. We restore the dependence on $\mu_R$ in our analytic results as discussed in \cref{app:mu}.

\subsection{Structure of the \texorpdfstring{$\qqgga$}{two-quark-two-gluon} amplitude}
\label{sec:decomposition-2q2gA}
\begin{figure}[t!]
  \begin{center}
    \includegraphics[width=0.9\textwidth]{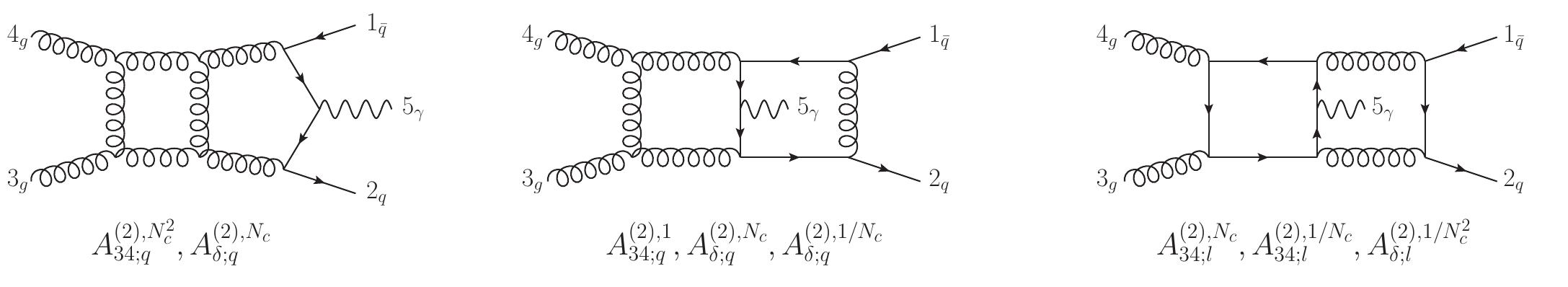}
  \end{center}
  \caption{Representative two-loop Feynman diagrams for the $\qqgga$ partonic process, 
           together with the partial amplitudes they contribute to.}
  \label{fig:qqgga2L}
\end{figure}

We label the scattering process as
\begin{equation}
    0 \rightarrow \bar{q}(p_1,h_1)+q(p_2,h_2)+g(p_3,h_3)+g(p_4,h_4)+\gamma(p_5,h_5) \,,
\label{eq:2q2gAlabel}
\end{equation}
where $h_i$ is the helicity of the particle with momentum $p_i$.
The $L$-loop amplitude $\cM^{(L)}$ has the following colour decomposition:
\begin{align}
    \begin{aligned}[b]
        \cM^{(L)}(1_{\bar{q}},2_{q},3_g,4_g,5_\gamma) &=  \sqrt{2} \, e \, g_s^2 \, n^L \, \bigg\lbrace
        (t^{a_3}t^{a_4})_{i_2}^{\;\;\bar i_1} \cA^{(L)}_{34}(1_{\bar{q}},2_q,3_g,4_g,5_\gamma) \\
        &+ (t^{a_4}t^{a_3})_{i_2}^{\;\;\bar i_1} \cA^{(L)}_{43}(1_{\bar{q}},2_q,3_g,4_g,5_\gamma)
        + \delta_{i_2}^{\;\;\bar i_1} \delta^{a_3 a_4} \cA^{(L)}_{\delta}(1_{\bar{q}},2_q,4_g,3_g,5_\gamma) \bigg\rbrace \,,
    \end{aligned}
\label{eq:2q2gAcolourdecomposition}
\end{align}
where $g_s$ and $e$ are the QCD and QED coupling constants, respectively, $n = m_\eps  \alpha_s/(4\pi)$, $\alpha_s = g_s^2/(4\pi)$, $m_\eps=\mathrm{i} (4\pi)^{\eps} e^{-\eps\gamma_E}$, and $t^a$ are the generators of $SU(N_c)$ in the fundamental representation,
normalised according to $\tr(t^a t^b) = \delta^{ab}/2$.
We note that
\begin{align}
    \cA^{(L)}_{43}(1_{\bar{q}},2_q,3_g,4_g,5_\gamma) &= \cA^{(L)}_{34}(1_{\bar{q}},2_q,4_g,3_g,5_\gamma) \,, &
    \cA^{(0)}_{\delta}(1_{\bar{q}},2_q,4_g,3_g,5_\gamma) &= 0 \,.
\end{align}
The partial amplitudes can be further decomposed into gauge invariant sub-amplitudes according to the closed fermion loop contribution, to the power of $N_c$, and to whether the photon is attached to an internal or external quark.
At one loop we have
\begin{subequations}
    \label{eq:2q2gAnfexpansion1L}
    \begin{align}
        \label{eq:2q2gAnfexpansion1Lt}
        \cA^{(1)}_{34} & = \chQ_q N_c A_{34;q}^{(1),N_c} + \chQ_q \frac{1}{N_c} A_{34;q}^{(1),1/N_c} + \chQ_q n_f A_{34;q}^{(1),n_f}
        + \bigg(\sum_l \chQ_l \bigg) A_{34;l}^{(1),1} \,, \\
        \label{eq:2q2gAnfexpansion1Ld}
        \cA^{(1)}_{\delta} & = \chQ_q A_{\delta;q}^{(1),1} + \bigg(\sum_l \chQ_l\bigg) \frac{1}{N_c} A_{\delta;l}^{(1),1/N_c} \,,
    \end{align}
\end{subequations}
while at two loops we have
\begin{subequations}
    \label{eq:2q2gAnfexpansion2L}
    \begin{align}
        \label{eq:2q2gAnfexpansion2Lt}
        \cA^{(2)}_{34} &= \chQ_q N_c^2 A_{34;q}^{(2),N_c^2} + \chQ_q A_{34;q}^{(2),1} +  \chQ_q \frac{1}{N_c^2} A_{34;q}^{(1),1/N_c^2}
        + \chQ_q N_c n_f A_{34;q}^{(2),N_c n_f} + \chQ_q\frac{n_f}{N_c}  A_{34;q}^{(2), n_f/N_c} \nn
        &+ \chQ_q n_f^2 A_{34;q}^{(2),n_f^2}
        + \bigg(\sum_l \chQ_l\bigg) N_c A_{34;l}^{(2), N_c} + \bigg(\sum_l \chQ_l\bigg) \frac{1}{N_c} A_{34;l}^{(2), 1/N_c} + \bigg(\sum_l \chQ_l\bigg) n_f  A_{34;l}^{(2), n_f} ,\\
        \label{eq:2q2gAnfexpansion2Ld}
        \cA^{(2)}_{\delta} &= \chQ_q N_c A_{\delta;q}^{(2),N_c} + \chQ_q \frac{1}{N_c} A_{\delta;q}^{(1),1/N_c}
        + \chQ_q n_f A_{\delta;q}^{(2),n_f} + \chQ_q \frac{n_f}{N_c^2}  A_{\delta;q}^{(2), n_f/N_c^2} \nn
        &+ \bigg(\sum_l \chQ_l\bigg) A_{\delta;l}^{(2), 1} + \bigg(\sum_l \chQ_l\bigg) \frac{1}{N_c^2} A_{\delta;l}^{(2), 1/N_c^2}
        + \bigg(\sum_l \chQ_l\bigg) \frac{n_f}{N_c} A_{\delta;l}^{(2), n_f/N_c} \,.
    \end{align}
\end{subequations}
We denote by $n_f$ the number of light quark flavours, and by $\chQ_{q}$ ($\chQ_l$) the charge of a flavour-$q$ ($l$) quark in units of $e$. 
We show representative Feynman diagrams contributing to the $\qqgga$ process at two loops in \cref{fig:qqgga2L}.
The right-hand sides of \cref{eq:2q2gAnfexpansion1L,eq:2q2gAnfexpansion2L} involve the following classes of partial amplitudes:
\begin{itemize}

\item $A^{(L),j}_{i;q}$: partial amplitude where the photon is attached to the external quark line of flavour $q$ (the first and second diagram in \cref{fig:qqgga2L}).
This type of amplitude always comes with a factor of $\chQ_q$.

\item $A^{(L),j}_{i;l}$: partial amplitude where the photon is attached to quarks running in the loop (the last diagram in \cref{fig:qqgga2L}).
This type of amplitude always comes with a factor of $\sum_{l}\chQ_l$, where we sum over the internal quarks of flavour $l$.

\end{itemize}

We derive the analytic expressions of the $A^{(L),j}_{34;i}$ and $A^{(L),j}_{\delta;i}$ partial amplitudes for the following independent helicity configurations:
\begin{align}
    &\texttt{$-++++$}, & &\texttt{$-+++-$}, & &\texttt{$-++-+$}, & &\texttt{$-+-++$}.
\end{align}
The remaining helicity configurations and partial amplitudes can be obtained by performing parity transformation and/or by permuting the external momenta.

It is convenient to factor out a combination of spinor products such that the helicity amplitude is free from the spinor phase.
This is particularly important when we work with the rational parameterisation of the external kinematics in terms of momentum twistor variables discussed in \cref{sec:amplitude_computation}: conjugation and permutation of external momenta must in fact be performed on phase-free quantities only.
For the $\qqgga$ partonic process we choose the following spinor phase factors:
\begin{subequations}
  \begin{align}
    \Phi_{\qqggaX}^{-++++} & = \frac{\spA{1}{2} \spA{3}{1} \spB{2}{3}}{\spA{2}{3} \spA{3}{4} \spA{4}{5}\spA{5}{1}} \,,\\
    \Phi_{\qqggaX}^{-+++-} & = \frac{\spA{1}{5}^3 \spA{2}{5}}{\spA{1}{2}\spA{2}{3} \spA{3}{4} \spA{4}{5}\spA{5}{1}}\,, \\
    \Phi_{\qqggaX}^{-++-+} & = \frac{\spA{1}{4}^3 \spA{2}{4}}{\spA{1}{2}\spA{2}{3} \spA{3}{4} \spA{4}{5}\spA{5}{1}} \,,\\
    \Phi_{\qqggaX}^{-+-++} & = \frac{\spA{1}{3}^3}{\spA{1}{2} \spA{3}{4} \spA{4}{5}\spA{5}{1}} \,. 
  \end{align}
  \label{eq:2q2gAphase}%
\end{subequations}

To obtain the squared matrix element we construct the colour-summed $L_1$-loop amplitude interfered with the $L_2$-loop amplitude as follows
\begin{align}
\sum_{\mathrm{colour}} \cM^{(L_1)*} \cM^{(L_2)} & = 2 e^2 g_s^4 \, n^{L_1} n^{L_2} (N_c^2-1) \bigg\lbrace 
  \frac{(N_c^2-1)}{4 N_c} \left[ \cA_{34}^{(L_1)*}\cA_{34}^{(L_2)} + \cA_{43}^{(L_1)*}\cA_{43}^{(L_2)} \right] \nn
 & \quad - \frac{1}{4 N_c} \left[ \cA_{34}^{(L_1)*}\cA_{43}^{(L_2)} + \cA_{43}^{(L_1)*}\cA_{34}^{(L_2)} \right] 
         + N_c \; \cA_{\delta}^{(L_1)*}\cA_{\delta}^{(L_2)} \label{eq:qqgga_squared} \\
 & \quad + \frac{1}{2} \left[  \cA_{34}^{(L_1)*}\cA_{\delta}^{(L_2)} + \cA_{43}^{(L_1)*}\cA_{\delta}^{(L_2)} 
                             + \cA_{\delta}^{(L_1)*}\cA_{34}^{(L_2)} + \cA_{\delta}^{(L_1)*}\cA_{43}^{(L_2)} \right] \bigg\rbrace \,. \nonumber
\end{align}

The bare $L$-loop amplitude $\cM^{(L)}$ contains both ultraviolet (UV) and infrared (IR) singularities.
The UV singularities are removed by means of renormalisation in the $\overline{\mathrm{MS}}$ scheme, while the IR ones can be predicted from the 
universal IR behaviour of QCD amplitudes~\cite{Catani:1998bh,Becher:2009qa,Becher:2009cu,Gardi:2009qi}.
We define the finite remainder of an amplitude, $\cR^{(L)}$, by subtracting the IR and UV singular parts from the bare amplitude as
\begin{subequations}
\label{eq:finrem}
  \begin{align}
    \label{eq:finrem1}
    |\cR^{(1)}\ra & = \left[ |\cM^{(1)}\ra - \frac{\beta_0}{\eps} |\cM^{(0)}\ra \right]  - \mathbf{Z}^{(1)} |\cM^{(0)}\ra \,, \\
    \label{eq:finrem2}
    |\cR^{(2)}\ra & = \left[ |\cM^{(2)}\ra - 2 \, \frac{\beta_0}{\eps} |\cM^{(1)}\ra - \left( \frac{\beta_1}{2\eps} - \frac{\beta_0^2}{\eps^2}  \right) |\cM^{(0)}\ra \right] - \mathbf{Z}^{(2)} |\cM^{(0)}\ra - \mathbf{Z}^{(1)} |\cR^{(1)}\ra \,,
  \end{align}
\end{subequations}
where $|\cM^{(L)}\ra$ is a vector in colour space built out of the partial amplitudes given in \cref{eq:2q2gAcolourdecomposition}. The square brackets on the right-hand sides of \cref{eq:finrem1,eq:finrem2} separate the subtraction of the UV poles from that of the IR ones.
The IR pole operator $\textbf{Z}$,
\begin{align}
\textbf{Z} = \textbf{I} + \frac{\alpha_s}{4\pi} \, \textbf{Z}^{(1)} + \left(\frac{\alpha_s}{4\pi}\right)^{2} \textbf{Z}^{(2)} + \mathcal{O}\left(\alpha_s^3\right) \,,
\end{align}
where $\textbf{I}$ is the identity, is a process-dependent matrix in colour space that encodes the IR singularities of the amplitude. We adopt the IR subtraction scheme of Refs.~\cite{Becher:2009cu,Becher:2009qa}; the $\mathbf{Z}^{(L)}$'s can be read off from Eq.~(12) of Ref.~\cite{Becher:2009qa}.
We give explicit expressions of $\mathbf{Z}^{(1)}$ and $\mathbf{Z}^{(2)}$ in the ancillary files~\cite{ppajjampsite}.
The colour decomposition of the colour-dressed finite remainder $\cR^{(L)}$ in terms of partial finite remainders $\cF^{(L)}$ for the $\qqgga$ partonic process is obtained from \cref{eq:2q2gAcolourdecomposition} by substituting
\begin{equation}
\cM^{(L)} \to \cR^{(L)} \qquad \mathrm{and} \qquad  \cA^{(L)}_{i} \to \cF^{(L)}_{i} \,.
\label{eq:finremreplacements1}
\end{equation}
Similarly, the $(N_c,n_f)$-decomposition of the $\qqgga$ partial finite remainders is derived from the bare amplitudes decomposition specified 
in \cref{eq:2q2gAnfexpansion1L,eq:2q2gAnfexpansion2L} by replacing
\begin{equation}
\cA^{(L)}_{i} \to \cF^{(L)}_{i} \qquad \mathrm{and} \qquad A^{(L),k}_{i;j} \to F^{(L),k}_{i;j} \,.
\label{eq:finremreplacements2}
\end{equation}

\subsection{Structure of the \texorpdfstring{$\qqQQa$}{four-quark} amplitude}
\label{sec:decomposition-2q2QA}

\begin{figure}[t!]
  \begin{center}
    \includegraphics[width=0.9\textwidth]{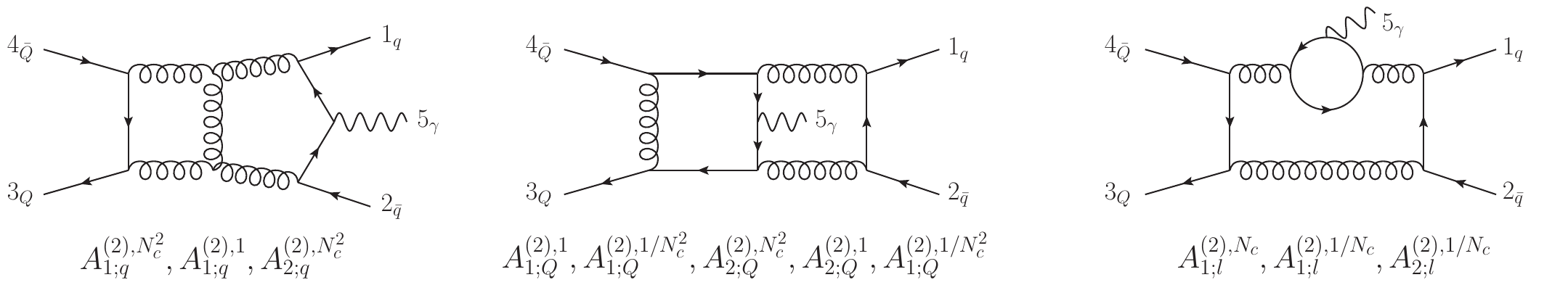}
  \end{center}
  \caption{Representative two-loop Feynman diagrams for the $\qqQQa$ partonic process,
           together with the partial amplitudes they contribute to.}
  \label{fig:qqQQa2L}
\end{figure}

For the 4-quark channel, we label the scattering process as follows,
\begin{equation}
    0 \rightarrow q(p_1,h_1) + \bar{q}(p_2,h_2) + Q(p_3,h_3) + \bar{Q}(p_4,h_4) + \gamma(p_5,h_5) \,, 
\label{eq:2q2QAlabel}
\end{equation}
where $q$ and $Q$ are massless quarks of distinct flavour. The identical-flavour case is discussed below.
Representative Feynman diagrams contributing to the $\qqQQa$ process at two loops are shown in \cref{fig:qqQQa2L}.
The colour decomposition of the $L$-loop amplitude $\cM^{(L)}$ is
\begin{align}
    \cM^{(L)}(1_q,2_{\bar{q}},3_Q,4_{\bar{Q}},5_\gamma) & =  \sqrt{2} \, e \, g_s^2 \, n^L \, \bigg\lbrace
    \delta_{i_1}^{\;\;\bar i_4} \delta_{i_3}^{\;\;\bar i_2} \cA^{(L)}_{1}(1_q,2_{\bar{q}},3_Q,4_{\bar{Q}},5_\gamma) \nn
    & \phantom{= \ } + \frac{1}{N_c}\delta_{i_1}^{\;\;\bar i_2} \delta_{i_3}^{\;\;\bar i_4} \cA^{(L)}_{2}(1_q,2_{\bar{q}},3_Q,4_{\bar{Q}},5_\gamma) \bigg\rbrace \,.
\label{eq:2q2QAcolourdecomposition}
\end{align}
We further decompose the partial amplitudes $\cA^{(L)}_{i}$ according the source of photon radiation, 
\begin{align}
    \begin{aligned}[b]
        \cA^{(L)}_{i}(1_q,2_{\bar{q}},3_Q,4_{\bar{Q}},5_\gamma) = \ &
        \chQ_q \, \cA^{(L)}_{i;q}(1_q,2_{\bar{q}},3_Q,4_{\bar{Q}},5_\gamma)
        + \chQ_Q \, \cA^{(L)}_{i;Q}(1_q,2_{\bar{q}},3_Q,4_{\bar{Q}},5_\gamma) \\
        & + \bigg(\sum_l \chQ_l\bigg) \cA^{(L)}_{i;l}(1_q,2_{\bar{q}},3_Q,4_{\bar{Q}},5_\gamma) \,,
    \end{aligned}
\end{align}
where we introduced the following sub-amplitudes:
\begin{itemize}
\item $\cA^{(L)}_{i;q}$: the photon is attached to the external $q$ quark line (the first diagram in \cref{fig:qqQQa2L});
\item $\cA^{(L)}_{i;Q}$: the photon is attached to the external $Q$ quark line (the second diagram in \cref{fig:qqQQa2L});
\item $\cA^{(L)}_{i,l}$: the photon is attached to an internal quark line (the third diagram in \cref{fig:qqQQa2L}).
      This class of sub-amplitudes vanishes both at tree level and one loop for the $\qqQQa$ process.
\end{itemize}
Furthermore, the $\cA^{(L)}_{i;Q}$ sub-amplitude can be obtained from the $\cA^{(L)}_{i;q}$ sub-amplitude via
\begin{align}
    \cA^{(L)}_{i;Q}(1_q,2_{\bar{q}},3_Q,4_{\bar{Q}},5_\gamma) &= \cA^{(L)}_{i;q}(3_Q,4_{\bar{Q}},1_q,2_{\bar{q}},5_\gamma), & i&=1,2\;.
\end{align}

In the case where the flavours of the scattering quark pairs are identical ($Q=q$),
\begin{equation}
    0 \rightarrow q(p_1,h_1) + \bar{q}(p_2,h_2) + q(p_3,h_3) + \bar{q}(p_4,h_4) + \gamma(p_5,h_5) \,, \nonumber
\end{equation}
the $L$-loop colour decomposition is
\begin{align}
    \cM^{(L)}(1_q,2_{\bar{q}},3_q,4_{\bar{q}},5_\gamma) = \ &  \sqrt{2} \, e \, g_s^2 \, n^L \, \bigg\lbrace 
       \delta_{i_1}^{\;\;\bar i_4} \delta_{i_3}^{\;\;\bar i_2} \cA^{(L)}_{1}(1_q,2_{\bar{q}},3_Q,4_{\bar{Q}},5_\gamma) \nn
    & + \frac{1}{N_c}\delta_{i_1}^{\;\;\bar i_2} \delta_{i_3}^{\;\;\bar i_4} \cA^{(L)}_{2}(1_q,2_{\bar{q}},3_Q,4_{\bar{Q}},5_\gamma) \nn
    &  +              \delta_{i_1}^{\;\;\bar i_2} \delta_{i_3}^{\;\;\bar i_4} \cA^{(L)}_{3}(1_q,2_{\bar{q}},3_Q,4_{\bar{Q}},5_\gamma) \nn
    & + \frac{1}{N_c}\delta_{i_1}^{\;\;\bar i_4} \delta_{i_3}^{\;\;\bar i_2} \cA^{(L)}_{4}(1_q,2_{\bar{q}},3_Q,4_{\bar{Q}},5_\gamma)
    \bigg\rbrace \,.
\end{align}
The $\cA^{(L)}_{3}$ and $\cA^{(L)}_{4}$ partial amplitudes are derived from $\cA^{(L)}_{1}$ and $\cA^{(L)}_{2}$ via
\begin{subequations}
    \begin{align}
        \cA^{(L)}_{3}(1_q,2_{\bar{q}},3_Q,4_{\bar{Q}},5_\gamma) = -\cA^{(L)}_{1}(1_q,4_{\bar{Q}},3_Q,2_{\bar{q}},5_\gamma) \,, \\
        \cA^{(L)}_{4}(1_q,2_{\bar{q}},3_Q,4_{\bar{Q}},5_\gamma) = -\cA^{(L)}_{2}(1_q,4_{\bar{Q}},3_Q,2_{\bar{q}},5_\gamma) \,.
    \end{align}
\end{subequations}

We further decompose the partial sub-amplitudes $\cA^{(L)}_{i;q}$ and $\cA^{(L)}_{i;l}$ ($i=1,\dots,4$) according to the closed fermion loop contribution and factor of $N_c$.
The $(N_c,n_f)$ decomposition at one loop is
\begin{equation}
    \cA^{(1)}_{i;q} = N_c A_{i;q}^{(1),N_c} + \frac{1}{N_c} A_{i;q}^{(1),1/N_c} + n_f A_{i;q}^{(1),n_f} \,,
\label{eq:2q2QAnfexpansion1L}
\end{equation}
while at two loops we have
\begin{subequations}
    \begin{align}
        \cA^{(2)}_{i;q} & =  N_c^2 A_{i;q}^{(2),N_c^2} +  A_{i;q}^{(2),1} + \frac{1}{N_c^2} A_{i;q}^{(2),1/N_c^2}
        + N_c \, n_f A_{i;q}^{(2),N_c n_f} + \frac{n_f}{N_c} A_{i;q}^{(2),n_f/N_c} + n_f^2 A_{i;q}^{(2),n_f^2} \,, \\
        \cA^{(2)}_{i;l} & = N_c \, A_{i;l}^{(2),N_c} + \frac{1}{N_c} A_{i;l}^{(2),1/N_c} \,.
    \end{align}
\label{eq:2q2QAnfexpansion2L}%
\end{subequations}
We derive the analytic expressions of the partial sub-amplitudes 
$A^{(L),i}_{1;q}$,
$A^{(L),i}_{2;q}$,
$A^{(L),i}_{1;l}$,
and
$A^{(L),i}_{2;l}$
for the following independent helicity configurations:
\begin{align}
    &\texttt{$-+-++$}\,, & &\texttt{$-++-+$}\,, & &\texttt{$+-+-+$}\,, & &\texttt{$+--++$}\,.
\end{align}
We obtain the remaining helicity configurations and partial amplitudes by performing conjugation and/or permuting the external momenta.
We choose the following spinor phase factors:
\begin{subequations}
  \begin{align}
    \Phi_{\qqQQaX}^{-+-++} & = \frac{\spA{1}{3}^3 \spA{2}{4}}{\spA{1}{2}\spA{2}{3} \spA{3}{4} \spA{4}{5}\spA{5}{1}} \,,\\
    \Phi_{\qqQQaX}^{-++-+} & = \frac{\spA{1}{4}^3 \spA{2}{3}}{\spA{1}{2}\spA{2}{3} \spA{3}{4} \spA{4}{5}\spA{5}{1}} \,,\\
    \Phi_{\qqQQaX}^{+-+-+} & = \frac{\spA{2}{4}^3 \spA{1}{3}}{\spA{1}{2}\spA{2}{3} \spA{3}{4} \spA{4}{5}\spA{5}{1}} \,,\\
    \Phi_{\qqQQaX}^{+--++} & = \frac{\spA{2}{3}^3 \spA{1}{4}}{\spA{1}{2}\spA{2}{3} \spA{3}{4} \spA{4}{5}\spA{5}{1}} \,.
  \end{align}
  \label{eq:2q2QAphase}%
\end{subequations}

We define the colour-summed $L_1$-loop amplitude interfered with the $L_2$-loop amplitude to build the squared matrix element for the $\qqqqa$ process as
\begin{align} \label{eq:qqqqa_squared}
\sum_{\mathrm{colour}} \cM^{(L_1)*} \cM^{(L_2)} = \ & 2 \, e^2 \, g_s^4 n^{L_1} n^{L_2}  \bigg\lbrace
          N_c^2 \left( \cA^{(L_1)*}_1 + \frac{1}{N_c} \cA^{(L_1)*}_4 \right) \left( \cA^{(L_2)}_1 + \frac{1}{N_c} \cA^{(L_2)}_4 \right) \nn
&  + N_c^2 \left( \cA^{(L_1)*}_3 + \frac{1}{N_c} \cA^{(L_1)*}_2 \right) \left( \cA^{(L_2)}_3 + \frac{1}{N_c} \cA^{(L_2)}_2 \right) \nn
&  + N_c \left( \cA^{(L_1)*}_1 + \frac{1}{N_c} \cA^{(L_1)*}_4 \right) \left( \cA^{(L_2)}_3 + \frac{1}{N_c} \cA^{(L_2)}_2 \right) \nn
&  + N_c \left( \cA^{(L_1)*}_3 + \frac{1}{N_c} \cA^{(L_1)*}_2 \right) \left( \cA^{(L_2)}_1 + \frac{1}{N_c} \cA^{(L_2)}_4 \right) \bigg\rbrace \,. 
\end{align}
To obtain the squared matrix element for $\qqQQa$ we set $\cA^{(L_2)}_3 = \cA^{(L_2)}_4 = 0$ in \cref{eq:qqqqa_squared}.

We define the finite remainders for the $\qqQQa$ partonic process according to \cref{eq:finrem}.
For $\qqQQa$, $|\cM^{(L)}\ra$ is a vector in colour space made of the partial amplitudes specified in \cref{eq:2q2QAcolourdecomposition}.
Similarly to the $\qqgga$ process, the colour decomposition of the finite remainder is obtained by applying to \cref{eq:2q2QAcolourdecomposition} the replacements in \cref{eq:finremreplacements1}, while the $(N_c,n_f)$-expansion of $\qqQQa$ partial finite remainders is derived by applying the substitutions specified in \cref{eq:finremreplacements2} to the bare amplitude decomposition in \cref{eq:2q2QAnfexpansion1L,eq:2q2QAnfexpansion2L}. 

\subsection{Helicity amplitude computation}
\label{sec:amplitude_computation}

We compute analytically the one- and two-loop helicity amplitudes contributing to $pp\to\gamma j j$ production using a workflow that
employs Feynman diagrams in conjunction with numerical sampling over finite fields and 
functional reconstruction techniques~\cite{vonManteuffel:2014ixa,Peraro:2016wsq} 
within the \textsc{FiniteFlow}~\cite{Peraro:2019svx} framework.
This approach has already been employed in the analytic computation of several 
two-loop amplitudes~\cite{Hartanto:2019uvl,Badger:2021owl,Badger:2021imn,Badger:2021ega}.

The computational tool-chain starts with the generation of Feynman diagrams using \textsc{Qgraf}~\cite{Nogueira:1991ex},  
followed by colour decomposition and filtering the partial amplitudes. We extract only the minimal set of independent partial amplitudes
to be further processed. We then construct the numerators of the loop amplitudes for all the independent helicity configurations.
These loop numerators are linear combinations of monomials involving loop-momentum dependent scalar products and spinor strings.
The coefficients of these monomials are functions of spinor products of external momenta. In order to perform the computation within
a finite field framework, we need to use a rational parametrisation of the external kinematics. 
To this end, we use the following momentum-twistor variables~\cite{Hodges:2009hk,Badger:2013gxa},
\begin{equation}
x_1  = s_{12}, \qquad
x_2  = -\frac{\trp(1234)}{s_{12}s_{34}}, \qquad
x_3  = -\frac{\trp(1345)}{s_{13}s_{45}}, \qquad
x_4  = \frac{s_{23}}{s_{12}} ,\qquad
x_5  = \frac{s_{45}}{s_{12}} ,
\label{eq:momtwistor}
\end{equation}
where $\trp(ijkl) = \tr((1+\gamma_5) \slashed{p}_i \slashed{p}_j \slashed{p}_k \slashed{p}_l)/2 $. We denote them cumulatively by $\vec{x} = \{x_1,\ldots,x_5\}$.
The momentum-twistor parametrisation in \cref{eq:momtwistor} may be used for the other massless five-point processes as well ($pp\to jjj$, $pp\to\gamma\gamma j$ and $pp\to\gamma\gamma\gamma$).
The only dimensionful variable in this parameterisation is $x_1$. We can thus set $x_1=1$, and recover its dependence at the end of the calculation.
The diagram processing and filtering, as well as the construction of helicity-dependent loop numerators, are done
using \textsc{Mathematica} and \textsc{Form}~\cite{Kuipers:2012rf,Ruijl:2017dtg} scripts, 
together with the library \textsc{Spinney}~\cite{Cullen:2010jv} 
to carry out the Dirac algebra and spinor manipulations.
We present our results in the 't~Hooft-Veltman scheme. 

We introduce the top-level integral families, which are a set of Feynman integral families that have the maximum number of loop propagators 
(in the two-loop five-point case there are at most 8 propagators, with 3 irreducible scalar products). The loop-momentum dependent terms in the loop numerators can then be written in terms of propagators of the top-level integral families, allowing us to express the loop scattering amplitude
as a linear combination of scalar Feynman integrals. The coefficients of these scalar Feynman integrals depend only on external kinematics in the
form of momentum twistor variables. 

We then reduce the scalar Feynman integrals to a set of pure~\cite{Henn:2013pwa} master integrals~\cite{Papadopoulos:2015jft,Gehrmann:2018yef,Abreu:2018rcw,Chicherin:2018mue,
Chicherin:2018old} by solving the integration-by-parts relations~\cite{Tkachov:1981wb,Chetyrkin:1981qh}. 
We generate the IBP relations among the integrals of the top-level families for a single ordering of the external legs, and solve them numerically over finite fields. We obtain the reduction of the integrals with other orderings of the external legs by evaluating the IBP solution at permuted points.

For all families, we generated identities among integrals without higher powers of propagators~\cite{Gluza:2010ws,Ita:2015tya,Larsen:2015ped} with respect to those appearing in the amplitude. We obtained them by using the Baikov representation of Feynman integrals~\cite{Baikov:1996rk,Baikov:1996cd}, where IBP identities generally contain both higher powers of propagators and dimensionally-shifted integrals. The latter can be eliminated by solving polynomial equations called syzygy equations. Equations without dimension-shifted integrals can be obtained as closed-form solutions, as illustrated in Ref.~\cite{Bohm:2017qme}. Following Ref.~\cite{vonManteuffel:2019wbj}, we put these solutions into a sparse matrix, and eliminated the higher powers of denominators through Gaussian elimination. We then reconstructed new template equations using \textsc{FiniteFlow}'s sparse solver~\cite{Peraro:2019svx}. From these, we generated a significantly smaller system of equations, which we then solved similarly to the more traditional Laporta algorithm~\cite{Laporta:2001dd}.

Next, we Laurent expand around $\eps=0$ up to the required order. The Laurent expansion of the master integrals is expressed in terms of a basis of special functions called pentagon functions and transcendental constants (e.g.\ $\pi, \zeta_3$...)~\cite{Chicherin:2020oor}. We denote the pentagon functions cumulatively by $f$, and the associated transcendental constants by $\mathfrak{c}$.
At this stage, we write the partial amplitudes as
\begin{equation}
A^{(L),k}_{i;j} =  \sum_{s=-2 L}^{o(L)} \sum_{r} \eps^s \, c_{r,s}(\vec{x}) \, \mathrm{mon}_r(f, \mathfrak{c}) \,,
\label{eq:dsdecomposition}
\end{equation}
where $o(1)=2$ and $o(2)=0$, and $\mathrm{mon}_r(f,\mathfrak{c})$ are monomials of pentagon functions $f$ and transcendental constants $\mathfrak{c}$.

In previous work~\cite{Badger:2021imn}, it was found that directly reconstructing the finite remainder decreased the reconstruction time, as it had a simpler expression than the full amplitude.
For $pp\to\gamma j j$ production, we find that the drop in polynomial degrees from amplitude to finite remainder is insignificant with respect to the reconstruction time. Therefore, we reconstruct the bare amplitude and subsequently compute the finite remainder.

In order to optimise the rational reconstruction of the amplitudes, we follow the strategy of Ref.~\cite{Badger:2021imn}. We refer to that work for a detailed discussion, and give here just an outline of the steps.
\begin{description}
\item[original{\normalfont:}] Reconstruction of the coefficients $c_{r,s}(\vec{x})$ in \cref{eq:dsdecomposition} without any optimisations.
\item[stage 1{\normalfont:}] We fit the $\mathbb{Q}$-linear relations among the coefficients $c_{r,s}(\vec{x})$, and solve them for the linearly independent coefficients, 
which are chosen to have the lowest polynomial degrees.
\item[stage 2{\normalfont:}] We determine the denominators of the coefficients $c_{r,s}(\vec{x})$ by reconstructing them analytically on a univariate slice~\cite{Abreu:2018zmy}, and matching the result against an ansatz. The latter is a product of letters of the pentagon alphabet~\cite{Chicherin:2017dob}, and spinor products ($\spA{i}{j}$, $\spB{i}{j}$). We then multiply away the identified factors.
\item[stage 3{\normalfont:}] We perform a univariate partial fraction decomposition of the coefficients $c_{r,s}(\vec{x})$ with respect to $x_4$ on the fly, and reconstruct the coefficients of the decomposition. The latter depend on $(x_2,x_3,x_5)$ only.
\item[stage 4{\normalfont:}] We perform another factor-matching (stage 2), this time on the coefficients of the partial fraction decompositions of the $c_{r,s}(\vec{x})$'s. We enlarge the ansatz used in stage 2 to include also the spurious factors introduced by the univariate partial fraction decomposition (stage 3).
\end{description}
Information on the degrees at each stage of the reconstruction for the most complicated reconstructions is shown in \cref{tab:degrees-2q2gA} for $\qqgga$ and in \cref{tab:degrees-2q2QA} for $\qqQQa$.
We provide the complete analytic expressions for the one- and two-loop finite remainders and IR/UV poles in the ancillary files~\cite{ppajjampsite}.

\renewcommand{\arraystretch}{1.5}
\begin{table}
    \centering
    \begin{tabular}{lllllll}
        \hline
        amplitude & helicity & original & stage 1 & stage 2 & stage 3 & stage 4 \\
        \hline
        $A^{(2),1}_{34;q}$ & $-++-+$ & 94/91 & 74/71 & 74/0 & 22/18 & 22/0 \\
        $A^{(2),1}_{34;q}$ & $-+-++$ & 93/89 & 90/86 & 90/0 & 24/14 & 18/0 \\
        $A^{(2),1/N_c^2}_{34;q}$ & $-++-+$ & 90/88 & 73/71 & 73/0 & 23/18 & 22/0 \\
        $A^{(2),1/N_c^2}_{34;q}$ & $-+-++$ & 90/86 & 86/82 & 86/0 & 24/14 & 19/0 \\
        $A^{(2),1/N_c}_{34;l}$ & $-+-++$ & 89/82 & 74/67 & 73/0 & 27/14 & 20/0 \\
        $A^{(2),1/N_c}_{34;l}$ & $-++-+$ & 85/81 & 61/58 & 60/0 & 27/18 & 20/0 \\
        \hline
        $A^{(2),N_c^2}_{34;q}$ & $-+-++$ & 58/55 & 54/51 & 53/0 & 20/16 & 20/0 \\
        \hline
    \end{tabular}
    \caption{
        Maximal numerator/denominator polynomial degrees of the rational coefficients of the six most complicated $\qqgga$ amplitudes ordered by the maximal numerator degree prior to any optimisation.
        The most complicated leading-colour contribution is also included in the last row of the table for comparison.
    }
    \label{tab:degrees-2q2gA}
\end{table}

\renewcommand{\arraystretch}{1.5}
\begin{table}
    \centering
    \begin{tabular}{lllllll}
        \hline
        amplitude & helicity & original & stage 1 & stage 2 & stage 3 & stage 4 \\
        \hline
        $A^{(2),1}_{2;q}$ & $-++-+$ & 83/80 & 53/50 & 53/0 & 20/14 & 16/0 \\
        $A^{(2),1/N_c^2}_{2;q}$ & $-++-+$ & 83/80 & 55/52 & 55/0 & 22/14 & 16/0 \\
        $A^{(2),1/N_c^2}_{2;q}$ & $-+-++$ & 83/80 & 62/59 & 62/0 & 22/13 & 17/0 \\
        $A^{(2),1}_{1;q}$ & $-+-++$ & 83/80 & 29/26 & 27/0 & 19/10 & 14/0 \\
        $A^{(2),1/N_c^2}_{1;q}$ & $-++-+$ & 83/80 & 39/35 & 39/0 & 26/16 & 20/0 \\
        $A^{(2),1/N_c^2}_{1;q}$ & $-+-++$ & 83/80 & 42/37 & 42/0 & 26/14 & 20/0 \\
        \hline
        $A^{(2),N_c^2}_{1;q}$ & $-++-+$ & 45/42 & 55/52 & 55/0 & 22/14 & 16/0 \\
        \hline
    \end{tabular}
    \caption{
        Same as \cref{tab:degrees-2q2gA}, but for $\qqQQa$.
    }
    \label{tab:degrees-2q2QA}
\end{table}

\subsection{Permutation and conjugation of the pentagon functions}
\label{sec:pfuncs_permutations}
The analytic expressions of the partial amplitudes are written in terms of rational coefficients and pentagon functions. The latter can be evaluated numerically in the $s_{12}$ channel through the library \textsc{PentagonFunctions++}~\cite{Chicherin:2020oor}.
To obtain all the helicity configurations and the full set of partial amplitudes from the independent set, 
we need to permute the external momenta entering the rational coefficients as well as the pentagon functions. Such permutations
may require crossings of initial-state momenta to the final state, preventing the evaluation with \textsc{PentagonFunctions++}. 
To overcome this, we exploit the fact that the complete set of pentagon functions is closed under permutations of the external momenta. In other words, any permutation $\sigma$ of the pentagon functions $f(\vec{s})$ can be expressed in terms of pentagon functions in the original ordering of the momenta as
\begin{align} \label{eq:permPfuncs}
\sigma \circ f(\vec{s}) := f\left(\sigma \circ \vec{s} \right) = M^{(\sigma)} \cdot \text{mon}\left( f(\vec{s}), \mathfrak{c} \right) \,,
\end{align}
where $M^{(\sigma)}$ is a matrix of rational numbers, and $\text{mon}\left( f(\vec{s}), \mathfrak{c} \right)$ is the vector of all monomials in $f(\vec{s})$ and $\mathfrak{c}$ up to transcendental weight 4.
We computed the matrices $M^{(\sigma)}$ from the expressions of all permutations of the master integrals in terms of pentagon functions provided in Ref.~\cite{Chicherin:2020oor}.
Thanks to \cref{eq:permPfuncs}, it suffices to evaluate the pentagon functions in one phase-space point in the $s_{12}$ channel to obtain the values of all their permutations as well. This not only removes the problem of evaluating outside the $s_{12}$ channel, but also reduces to the minimum the number of calls to \textsc{PentagonFunctions++}. Since the latter are the most expensive part of the numerical evaluation of the two-loop amplitudes, this approach is also more efficient.

The pentagon functions (as well as the associated transcendental constants) have well-defined behaviour under parity conjugation. They are either even or odd, i.e., they either stay the same or change sign. Conjugating them thus amounts to flipping the sign of the odd ones. In this regard, note that the pseudo-scalar invariant $\tr_5$ is purely imaginary in any physical channel. The library \textsc{PentagonFunctions++} takes $\tr_5$ to have a positive imaginary part. The values of the pentagon functions for the opposite choice can be obtained by parity conjugation.

We provide the pentagon-function permutation rules required to obtain the full set of helicity configurations and partial amplitudes for 
$pp\to\gamma j j$ production in the ancillary files under the folder \verb=amplitudes/permrules=~\cite{ppajjampsite}.

\subsection{Validation}
We perform the following checks to validate our results for the amplitudes.

\begin{description}

\item[Ward-identity test.]
We verify the gauge invariance of the bare helicity amplitudes by 
replacing the gluon or photon polarisation vector by its momentum, and checking that the resulting amplitudes vanish.

\item[UV and IR poles subtraction.]
The fact that the finite remainders in \cref{eq:finrem} are indeed finite at $\eps=0$ proves that the amplitudes have the expected UV and IR poles.

\item[Comparison against \textsc{OpenLoops}.]
We compare our results numerically against \textsc{OpenLoops} at the level of the squared matrix element for the tree-level amplitudes 
and the one-loop finite remainders (at $\cO(\eps^0)$), for all scattering channels contributing to $pp\to \gamma jj$ production.

\item[Checks on the symmetry of the amplitude.]
We perform several checks to demonstrate that the amplitudes have the expected symmetries. For example, the $\qqgga$ subprocess is symmetric under the exchanges $1 \to 2$ and $3 \to 4$. We apply the exchange in two ways: (1) by permuting rational coefficients and pentagon functions, and (2) by permuting the external momenta in the numerical input. 
We check that the two approaches give the same results.

\item[One-loop amplitude check through $\cO(\eps^2)$.]
To ensure that the $\tr_5$ invariant is correctly evaluated when performing permutations and conjugation as well as 
in the numerical evaluation, we derive the full set of partial amplitudes for all the contributing helicity configurations in terms of  
\textit{scalar} master integrals. Unlike some of the pure master integrals, the latter are parity even, i.e., they do not depend on the sign of $\tr_5$. We then evaluate the scalar master integrals numerically using the package \textsc{AMFlow}~\cite{Liu:2022chg} up to the necessary $\eps$ order such that the amplitude is evaluated through $\cO(\eps^2)$. The amplitude computed using this approach is then 
compared against the amplitude obtained by permutation and/or conjugation.

\end{description}

\subsection{Numerical evaluation}

We implemented the amplitudes presented above in a \textsc{C++} code that yields the hard functions, defined in \cref{app:benchmark}, by performing the colour and helicity sums for the IR- and UV-finite squared remainders according to Eqs.~\eqref{eq:qqgga_squared} and \eqref{eq:qqqqa_squared}, as required for the evaluation of double-virtual contributions to differential cross sections. The code utilises the numerical implementation of the special functions provided in \textsc{PentagonFunctions++}~\cite{Chicherin:2020oor}, and has been checked against an independent implementation in \textsc{Mathematica} for a number of phase-space points. If needed, higher-precision floating-point arithmetic is available through the \textsc{qd} library \cite{QDlib}, which implements the double-double and quad-double data types. This is indeed necessary, since cancellations between rational terms and/or special functions for some phase-space points lead to a loss of precision resulting in values without a sufficient number of correct digits. In order to capture these problematic cases, we employ the following algorithm for each phase-space point $p = \{p_i\}_{i=1}^5$:
\begin{enumerate}
\item[1)] The two-loop hard function $\cH^{(2)}(p,\mu_R)$ is evaluated at $p$ with renormalisation scale $\mu_R$ in standard double precision;
\item[2)] Additionally, the two-loop hard function is evaluated at $p'= \{ a \times p_i \}_{i=1}^5$ and renormalisation scale $\mu^{\prime}_R = a \times \mu_R$, with $a$ a numerical constant. The quantity
  \begin{equation}
    r^{\text{diff}} = \frac{|\cH^{(2)}(p,\mu_R) - a^2 \, \cH^{(2)}(p', \mu^{\prime}_R)|}{|\cH^{(2)}(p,\mu_R) + a^2 \, \cH^{(2)}(p', \mu^{\prime}_R)|}\;,
  \end{equation}
  provides an estimate of the relative numerical precision of the squared matrix element. Here, we have taken into account the mass dimension of $\cH^{(2)}(p,\mu_R)$ for a $2 \to 3$ scattering process. The estimated ``number of correct digits'' $n^{\text{cor}}$ is given by
  \begin{equation}
    n^{\text{cor}} = -\log_{10}(r^{\text{diff}}) \;.
  \end{equation}
\item[3)] If $n^{\text{cor}} > 4$ the phase-space point is accepted and the algorithm restarts from 1) for the next phase-space point. If $n^{\text{cor}} \leq 4$ the floating point precision is increased by switching to the next larger data type and the algorithm restarts from 1). Instead of simply using the same data type for every operation during amplitude evaluation, we implement the following precision levels:
  \begin{description}
    \item[L1{\normalfont:}] pentagon and rational functions are evaluated with double precision ($\approx 16$ digits);
    \item[L2{\normalfont:}] pentagon functions are evaluated with double precision, while the rational functions with double-double precision ($\approx 32$ digits);
    \item[L3{\normalfont:}] pentagon functions are evaluated with double-double precision, while the rational functions with quad-double precision ($\approx 64$ digits).
  \end{description}
\end{enumerate}

\begin{figure}
    \centering
    \includegraphics{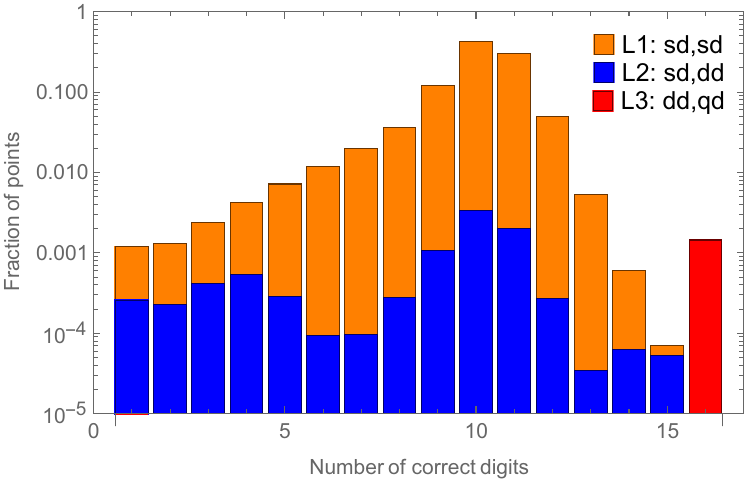}
    \caption{Fraction of phase-space points with a given number of correct digits with respect to the total number of points. The labels ``L1: sd,sd'', ``L2: sd,dd'', and ``L3: dd,qd'' denote the implementation levels L1, L2 and L3. The labels ``sd'',  ``dd'', and ``qd'' refer to single-double, double-double, and quad-double precision. Further details are given in the text.}
    \label{fig:vvf_stats}
\end{figure}

In order to obtain the differential distributions presented in \cref{sec:phenomenology}, we have evaluated the squared matrix elements at $\approx 2.2$ million phase-space points. The distribution of the estimated numerical precision for the three different precision levels is shown in \cref{fig:vvf_stats}. About $1\%$ of the phase-space points required at least level L2 to achieve four correct digits. Among those points, $10\%$ were identified as still unstable and had to be evaluated at L3. About $0.1\%$ of the whole set of phase-space points required an evaluation at L3 to achieve the precision goal of four correct digits.

The total computing time for the double-virtual contributions was $10$ kCPUh corresponding roughly to $16$ seconds on average for the evaluation at a single phase-space point. This timing includes stability checks and potential re-evaluations with higher precision.

\section{LHC Phenomenology at 13 TeV}\label{sec:phenomenology}

\subsection{Setup}

The evaluation of perturbative second-order corrections to the differential cross sections for photon + di-jet production requires the combination of double-virtual, real-virtual and double-real contributions. To this end, we employ \textsc{Stripper}, an implementation of the sector-improved residue subtraction scheme \cite{Czakon:2010td, Czakon:2014oma, Czakon:2019tmo}, which also provides the cross sections up to next-to-leading order. This software has been used previously in various cross-section calculations and has been validated in numerous comparisons to public results. Tree-level matrix elements are implemented through the \textsc{AvH} library \cite{Bury:2015dla}. For the single-virtual contributions, which require one-loop matrix elements with one photon and up to three jets, we employ \textsc{OpenLoops2} \cite{Buccioni:2019sur}. The evaluation of the squared matrix elements needed for the double-virtual corrections has been described in the previous Section.

We work in the five-flavour, $n_f = 5$, scheme, i.e.\ we consider QCD with five massless quarks and ignore the existence of top-quarks. All predictions presented in this section use the \textsc{NNPDF31} \cite{NNPDF:2017mvq} PDF set as implemented in \textsc{LHAPDF} \cite{Buckley:2014ana}. We compute the cross section for two different nominal choices of the renormalization ($\mu_R$) and factorization scale ($\mu_F$):
\begin{subequations}
\begin{align}
    \mu_R &= \mu_F = H_T = E_{\perp}(\gamma) + p_T(j_1) + p_T(j_2)\;\quad\text{and}\\
    \mu_R &= \mu_F = E_{\perp}(\gamma)\;,
\end{align}
\end{subequations}
where $E_{\perp}(\gamma)$ is the transverse energy of the photon and $p_T(j_i)$ denotes the $i$-th leading jet transverse momentum. We estimate the uncertainty from missing higher orders by performing a conventional 7-point scale variation around the nominal $\mu_{R,F}$ values by factors of $2$ subject to the constraint $\frac{1}{2} \leq \frac{\mu_R}{\mu_F} \leq 2$.

In order to showcase the theoretical NNLO QCD predictions for differential cross sections, we will compare our results to measurement data obtained by the ATLAS collaboration at 13 TeV \cite{ATLAS:2019iaa}. As already explained in the introduction, the final-state photon may either be produced through hard emission, well separated from other partons, or through collinear emission described within the fragmentation formalism. Since the main achievement of the current publication is the evaluation of the two-loop virtual corrections for the two-to-three processes with four partons and one photon, our attention is devoted to the phase space region which is well described by fixed-order perturbative methods without collinear enhancements. Fortunately, the measurements presented in Ref.~\cite{ATLAS:2019iaa} are divided into two samples, one of which is more sensitive to the perturbative and one which is more sensitive to the fragmentation component. Of course, there is always some contamination in each region from effects characteristic of the other region. We shall estimate the size of the unwanted effects below, and argue that they are well within the uncertainties of the measurement data to which we compare.

Since we adopt the phase-space of Ref.~\cite{ATLAS:2019iaa}, our event-selection cuts are defined as follows:
\begin{enumerate}

    \item We require at least two jets defined with the anti-$k_T$ algorithm \cite{Cacciari:2008gp} for jet radius $R=0.4$ that have minimal transverse momentum of $p_T(j) > 100\;\text{GeV}$ and maximal rapidity $|\eta(j)| < 2.5$.

    \item The identified jets must be separated from the photon by $\Delta R (\gamma,j) > 0.8$.
    
    \item One isolated photon must be present in the final state with $E_{\perp}(\gamma) \geq 150 \;\text{GeV}$, $|\eta(\gamma)| \leq 2.37$ excluding $1.37 \leq |\eta(\gamma)| \leq 1.56 $.
    
    \item It is well-known that hard photon isolation criteria lead to infrared sensitivity beyond the Born approximation, which manifests at fixed perturbative order by uncompensated singularities. These singularities may be absorbed into a fragmentation function. However, there is a simpler approach that allows us to avoid the implementation of the fragmentation formalism. This approach is based on a smooth photon isolation criterion proposed by Frixione in Ref.~\cite{Frixione:1998jh}. Upon extension with an additional hard-cone isolation \cite{Siegert:2016bre, Chen:2022gpk}, this method simulates the experimental setup very closely. This hybrid isolation criterion has been adopted in Section 4.3 of Ref.~\cite{ATLAS:2019iaa} for the generation of next-to-leading order QCD predictions with the Monte-Carlo generator \textsc{Sherpa} \cite{Gleisberg:2008ta, Sherpa:2019gpd} used for comparisons to data in that publication. Accordingly, we accept an event if the sum, $E_{\perp}(r)$, of the transverse energies of all partons separated from the photon by the angular distance $\Delta R \leq r$ satisfies the smooth-cone (Frixione) condition:
    \begin{equation}
    E_{\perp}(r) \leq E_{\perp\rm{max}}(r) = 0.1 \, E_\perp(\gamma) \Bigg( \frac{1-\cos(r)}{1-\cos(R_{\rm max})} \Bigg)^2 \quad \text{for} \quad r \leq R_{\rm max} = 0.1 \; ,
    \end{equation}
    as well as the hard-cone condition:
    \begin{equation} \label{eq:hardcone}
    E_{\perp}(r) \leq E_{\perp\rm{max}} = 0.0042 \, E_{\perp}(\gamma) + 10\;\text{GeV} \quad \text{for} \quad r \leq R_{\rm max} = 0.4 \; .
    \end{equation}
\end{enumerate}

These event-selection cuts define the {\it inclusive sample} used in the determination of the differential cross sections. The aforementioned reduction of the fragmentation effects is further strengthened by imposing the requirement \cite{ATLAS:2019iaa}:
\begin{equation}
E_{\perp}(\gamma) > p_T(j_1) \; ,
\end{equation}
which defines the {\it direct-enriched} sample. The {\it fragmentation-enriched} sample has been defined in Ref.~\cite{ATLAS:2019iaa} by the requirement $E_{\perp}(\gamma) < p_T(j_2)$. Here, $j_1$ and $j_2$ are the hardest (leading) and next-to-hardest (sub-leading) jet ordered according to their transverse momenta.

\subsection{Selected observables}

In this section, we discuss selected results for the {\it inclusive} and {\it direct-enriched} phase space defined above. Results for the following distributions in all three phase spaces are provided in numeric format together with this publication:
\begin{enumerate}
    \item $E_{\perp}(\gamma)$: photon transverse energy,
    \item $p_T^{\text{jet}}$: jet transverse momentum,
    \item $y^{\text{jet}}$: jet pseudorapidity,
    \item $|\Delta y^{\gamma - \text{jet}}|$: absolute value of the pseudorapidity difference between the photon and the jet,
    \item $|\Delta \phi^{\gamma - \text{jet}}|$: azimuthal-angle difference between the photon and the jet,
    \item $|\Delta y^{j_1 - j_2}|$: absolute value of the pseudorapidity difference between the leading and sub-leading jet,
    \item $|\Delta \phi^{j_1 - j_2}|$: azimuthal-angle difference between the leading and sub-leading jet,
    \item $m(j_1j_2)$: invariant mass of the leading and sub-leading jet,
    \item $m(\gamma j_1 j_2)$: invariant mass of the photon, leading and sub-leading jet.
\end{enumerate}
In the case of $p_T^{\text{jet}}, y^{\text{jet}}, |\Delta y^{\gamma - \text{jet}}|$ and $|\Delta \phi^{\gamma - \text{jet}}|$, every event is counted twice by filling the given histogram using the value of the respective observable for both the leading and the sub-leading jet.

Together with the LO, NLO and NNLO QCD predictions, we show not only the measurement results from Ref.~\cite{ATLAS:2019iaa} but also \textsc{Sherpa} predictions provided by ATLAS. The latter are obtained from an NLO-matched QCD parton-shower merged with LO photon+four-jet samples corresponding to the double-real radiation contributions in the presented NNLO QCD predictions. The \textsc{Sherpa} predictions use $E_{\perp}(\gamma)$ for the renormalisation and factorisation scale. More details can be found in Ref.~\cite{ATLAS:2019iaa}.

\paragraph{Transverse photon energy $E_{\perp}(\gamma)$.}

\begin{figure}
    \centering
    \includegraphics[width=0.5\textwidth,page=1]{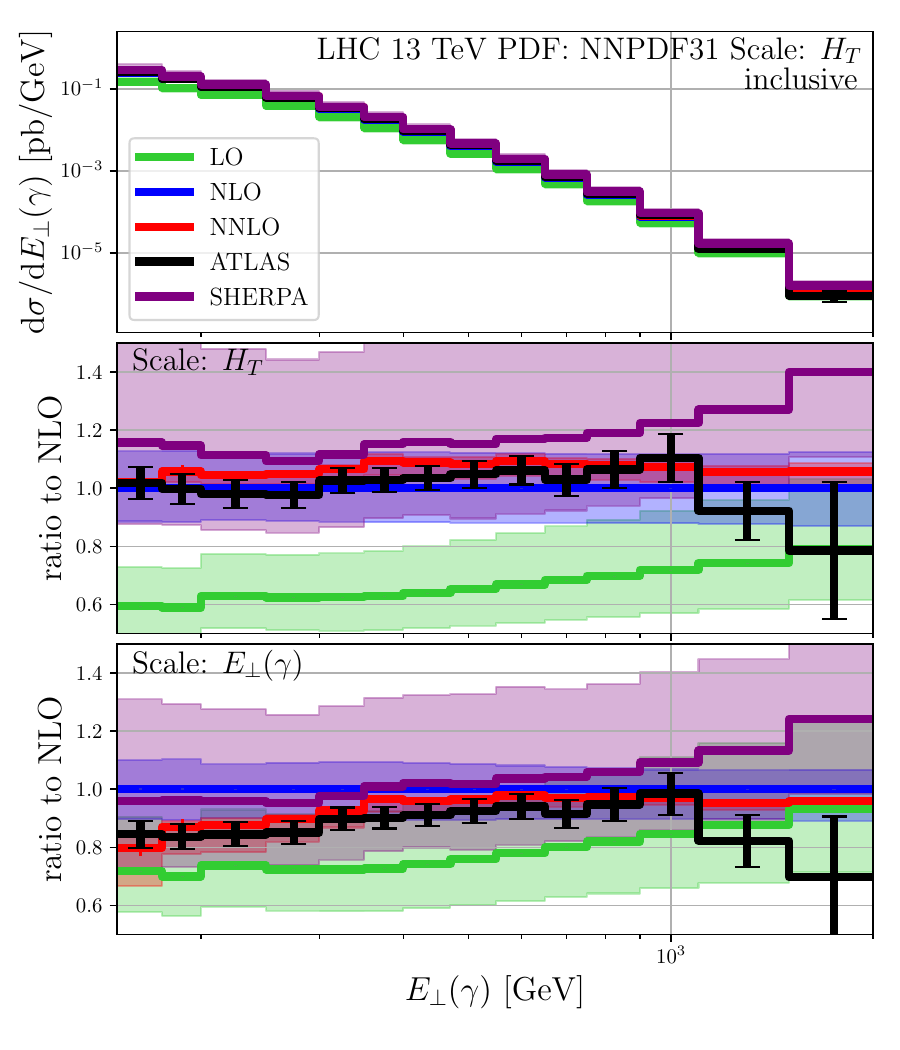}%
    \includegraphics[width=0.5\textwidth,page=1]{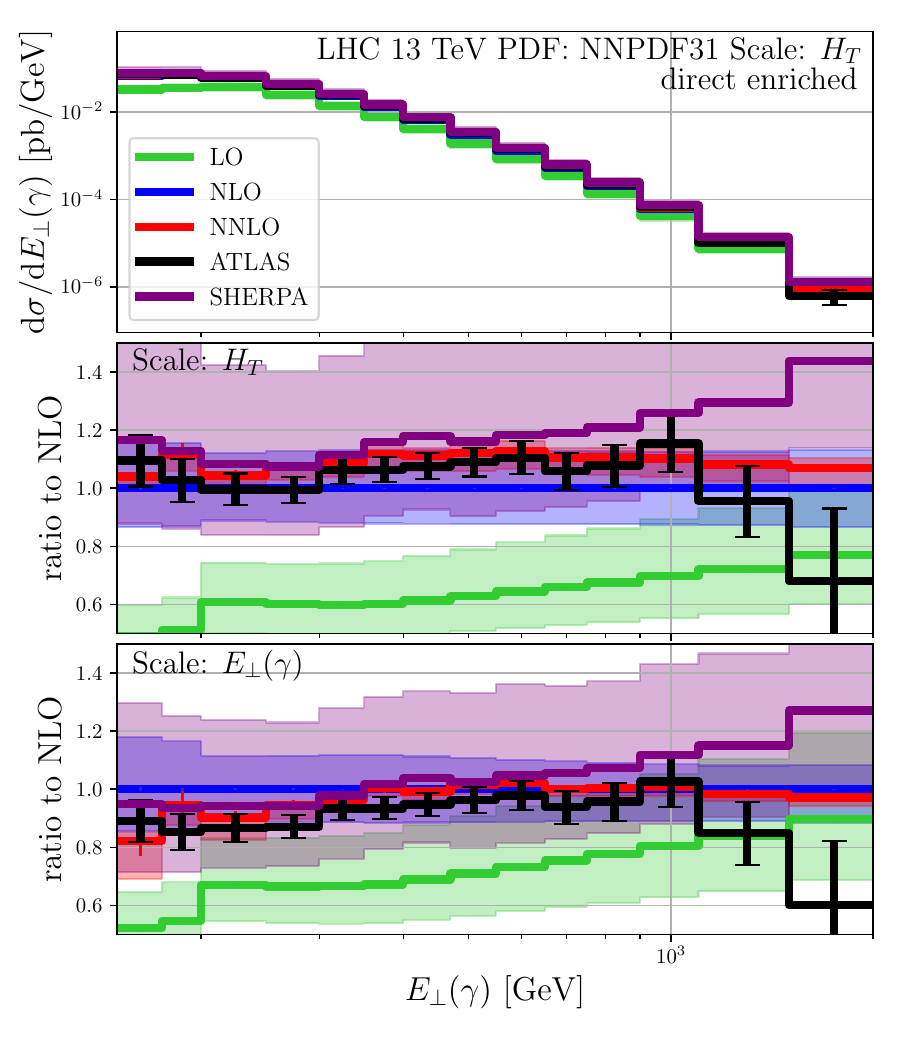}
    \caption{Differential cross sections w.r.t.\ the transverse energy of the photon $E_{\perp}(\gamma)$ in the {\it inclusive} (left plot) and {\it direct-enriched} (right plot) phase space at LO (green), NLO (blue) and NNLO (red) QCD compared to data (black) and \textsc{Sherpa} (purple) prediction provided by ATLAS\cite{ATLAS:2019iaa}. The top panels show the absolute values for the $H_T$ scale choice. The middle (bottom) panel shows the ratio to NLO QCD using the $H_T$ ($E_{\perp}(\gamma)$) scale. The coloured bands show scale variation and the vertical coloured bars indicate statistical uncertainties.}
    \label{fig:diffxs_ET}
\end{figure}

The differential cross section is shown in \cref{fig:diffxs_ET} for the {\it inclusive} (left plot) and for the {\it direct-enriched} phase space (right plot). For the $H_T$ scale choice, the NLO QCD corrections are about $80\%$ at small transverse energies and about $30\%$ at high energies relative to LO for both samples. The NLO QCD predictions have sizeable estimated uncertainties from missing higher orders of about $10-20\%$. Nevertheless, they describe the ATLAS data within those uncertainties. The NNLO QCD corrections are positive and much smaller than the NLO. They vary from $\approx 1\%$ at low transverse energy to maximally $10\%$ for $E_{\perp}(\gamma) \approx 400-600\;\text{GeV}$. The scale dependencies decrease significantly to $1-5\%$ when including the second-order corrections. Moreover, the NNLO QCD predictions describe the data better than the NLO as far as the distribution's shape is concerned. On the other hand, the normalisation is slightly high albeit still compatible within the systematic uncertainties. The picture changes when considering results obtained with the $E_{\perp}(\gamma)$ scale. The LO predictions are significantly larger than for the $H_T$ scale, while the NLO corrections are smaller. In this case, the NNLO QCD corrections are negative and sizeable (up to $-20\%$) for small transverse energies. The data is again well described although with larger remaining scale dependence. The large corrections and uncertainties indicate that the scale choice $E_{\perp}(\gamma)$ does not capture the relevant kinematic scales entering this observable. This is different from the inclusive photon production case $pp \to \gamma + X$. The reason lies most likely in the presence of new scales introduced through the jet-selection requirements.

At higher energies, while still agreeing with data within the uncertainties, the predicted spectrum is harder than the measurement. It has been shown in Refs.~\cite{Kuhn:2005gv, Becher:2013zua, Becher:2015yea} that electroweak (EW) corrections are significant in the TeV region, and can reach about $\approx -10\%$ for $E_{\perp}(\gamma) > 1 \; \text{TeV}$ in $pp\to\gamma j$. Similar corrections can be expected for $pp\to\gamma jj$ and could alleviate the observed discrepancy at high energies. The $E_{\perp}(\gamma)$ distribution is not the only one affected by potentially large EW corrections. In fact, for large invariant jet pair mass $m(j_1j_2)$ and large rapidity separation between the two jets, sizeable corrections are also expected. The inclusion of these corrections is, however, left for future work.

A striking observation is that the description of the data by the purely perturbative QCD (pQCD) calculation is noticeably better than by the \textsc{Sherpa} prediction which includes parton-shower effects and higher jet multiplicities through merging. It appears that the \textsc{Sherpa} predictions systematically overshoot the measurement at high energies, while pQCD predictions are closer to data both in size and in shape. The uncertainties of the \textsc{Sherpa} predictions estimated using scale variation are twice as large as those of the fixed order calculation. This is the reason why the \textsc{Sherpa} predictions still agree with data. A possible explanation for the size of the uncertainties despite the fact that both approaches have at least NLO accuracy could be the multi-jet merging with the photon+4-jet sample, which is likely to come with a large scale dependence.

\paragraph{Jet transverse momentum $p_T^{\text{jet}}$.}

\begin{figure}
    \centering
    \includegraphics[width=0.5\textwidth,page=2]{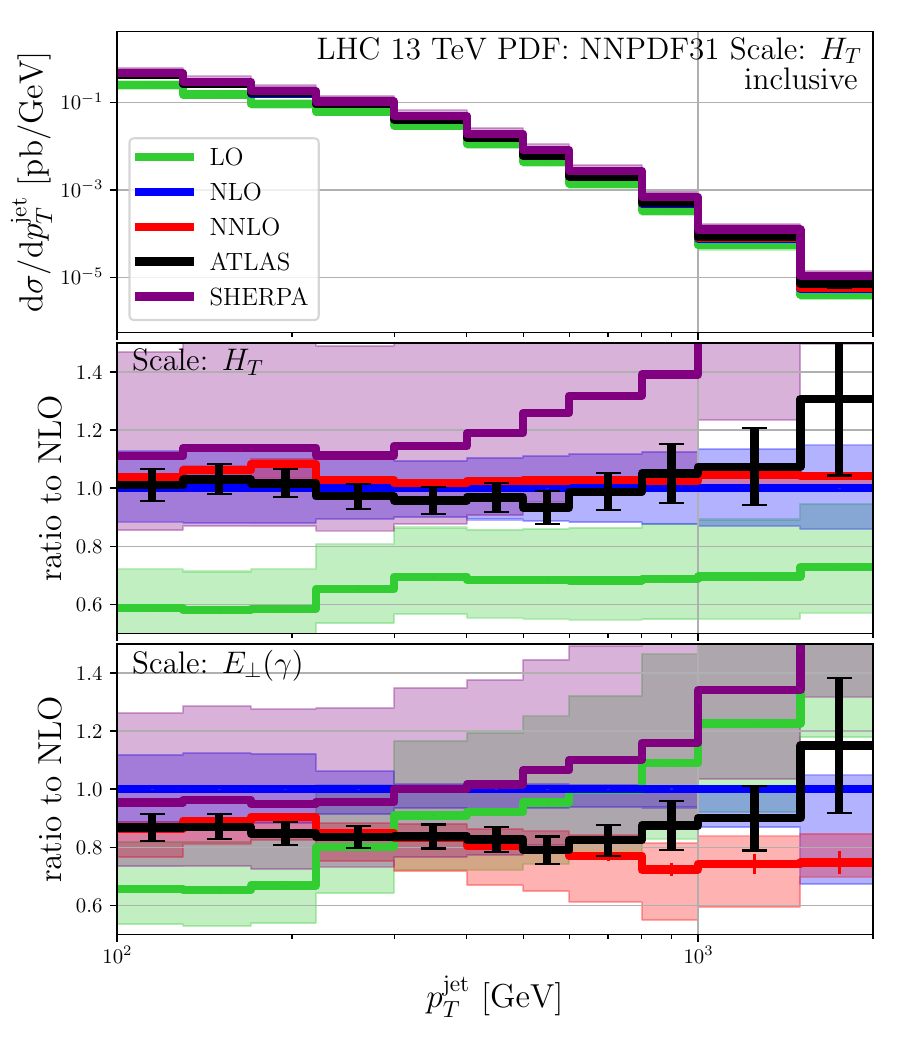}%
    \includegraphics[width=0.5\textwidth,page=2]{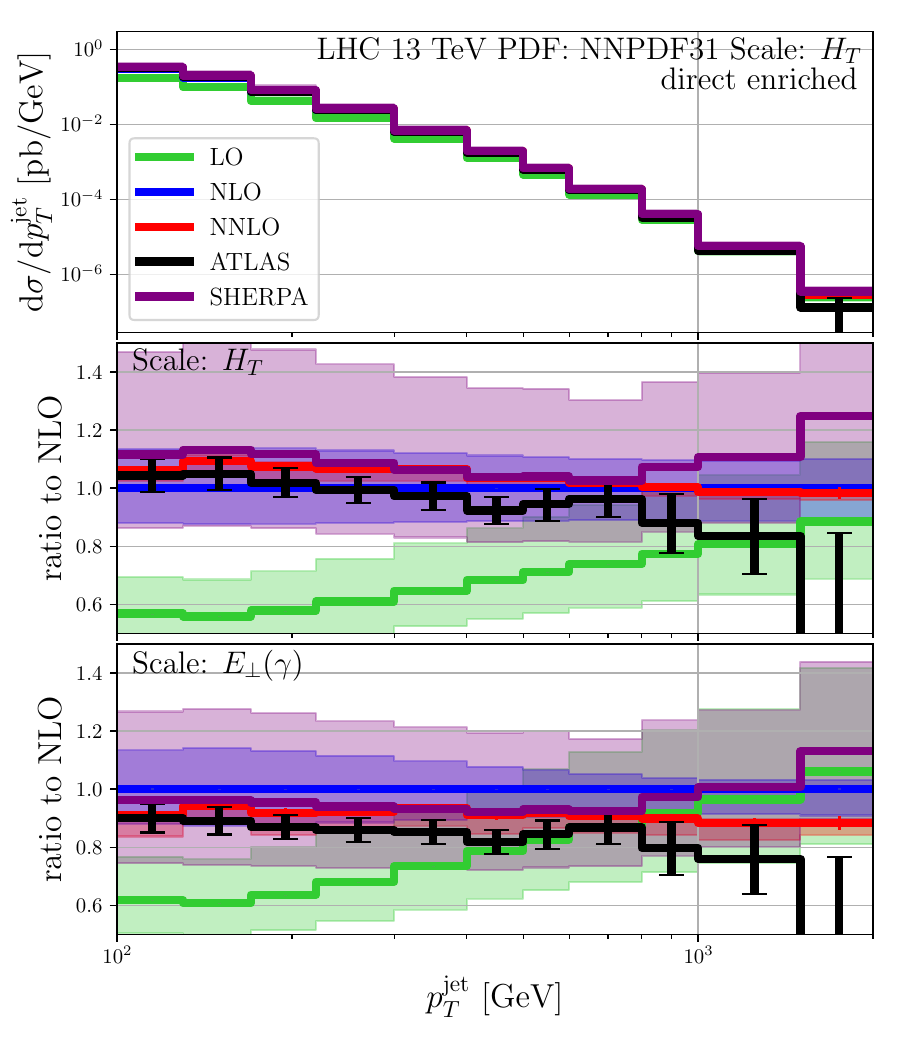}
    \caption{The same as \cref{fig:diffxs_ET} but for the $p_T^{\text{jet}}$ observable.}
    \label{fig:diffxs_pTj}
\end{figure}

In this case, we make similar observations as for the transverse energy of the photon. The differential cross section is presented in \cref{fig:diffxs_pTj}. For the $H_T$ scale, we find small NNLO QCD corrections and small remaining scale dependence. The description of the data is very clearly improved with respect to NLO QCD and \textsc{Sherpa}. The observable demonstrates once more the sub-optimal properties of the $E_{\perp}(\gamma)$ scale. Looking at the {\it inclusive} phase space, the NNLO QCD scale dependence is much larger than in the case of $H_T$, especially at high transverse momentum. This is not the case in the {\it direct-enriched} phase space, where both scales behave similarly. This is consistent with the fact that the two scale choices are closest (kinematics-wise per event) when $E_{\perp}(\gamma) > p_T(j_1)$ since in that case, $E_{\perp}(\gamma)$ dominates the value of $H_T$. This suggests that the region where $E_{\perp}(\gamma) < p_T(j_1)$ is particularly poorly described with the $E_{\perp}(\gamma)$ scale because this scale is relatively low (compared to the jet momenta), leading to comparatively large $\alpha_S(\mu_R)$ and thus large corrections.

\paragraph{Invariant mass of the jet pair $m(j_1j_2)$.}

\begin{figure}
    \centering
    \includegraphics[width=0.5\textwidth,page=8]{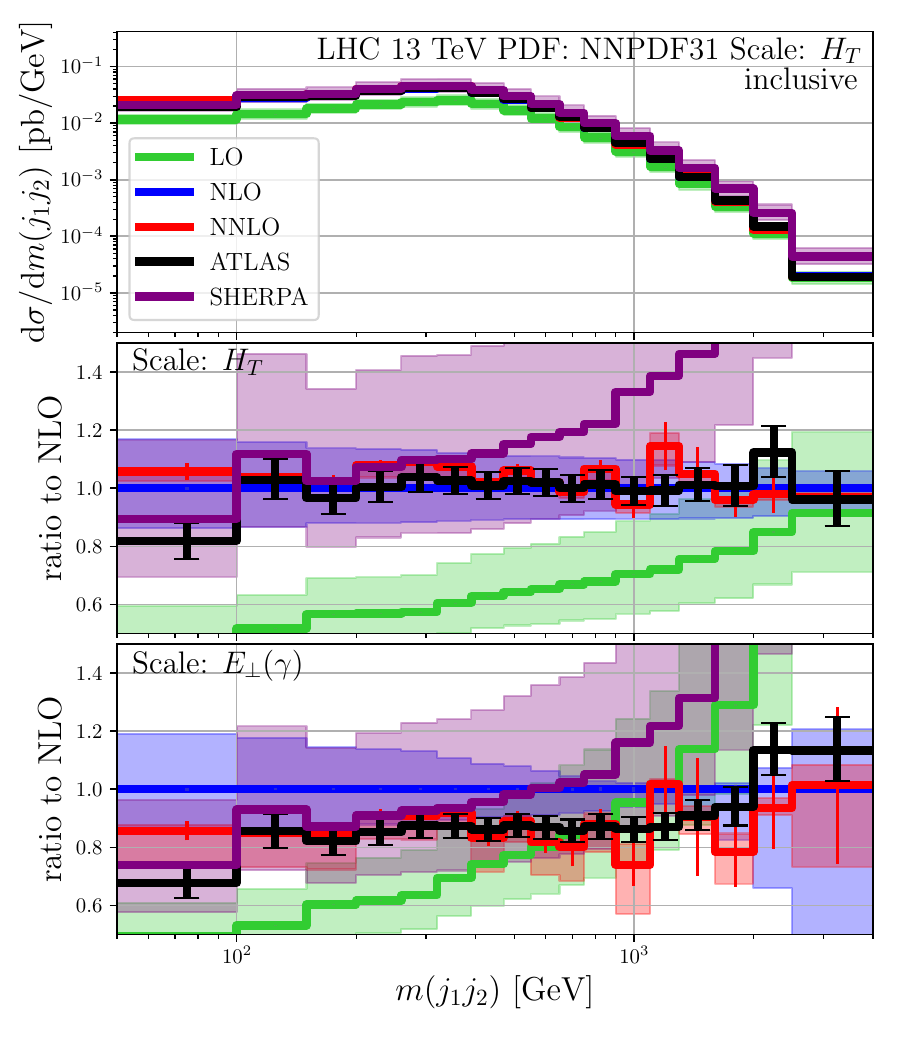}%
    \includegraphics[width=0.5\textwidth,page=8]{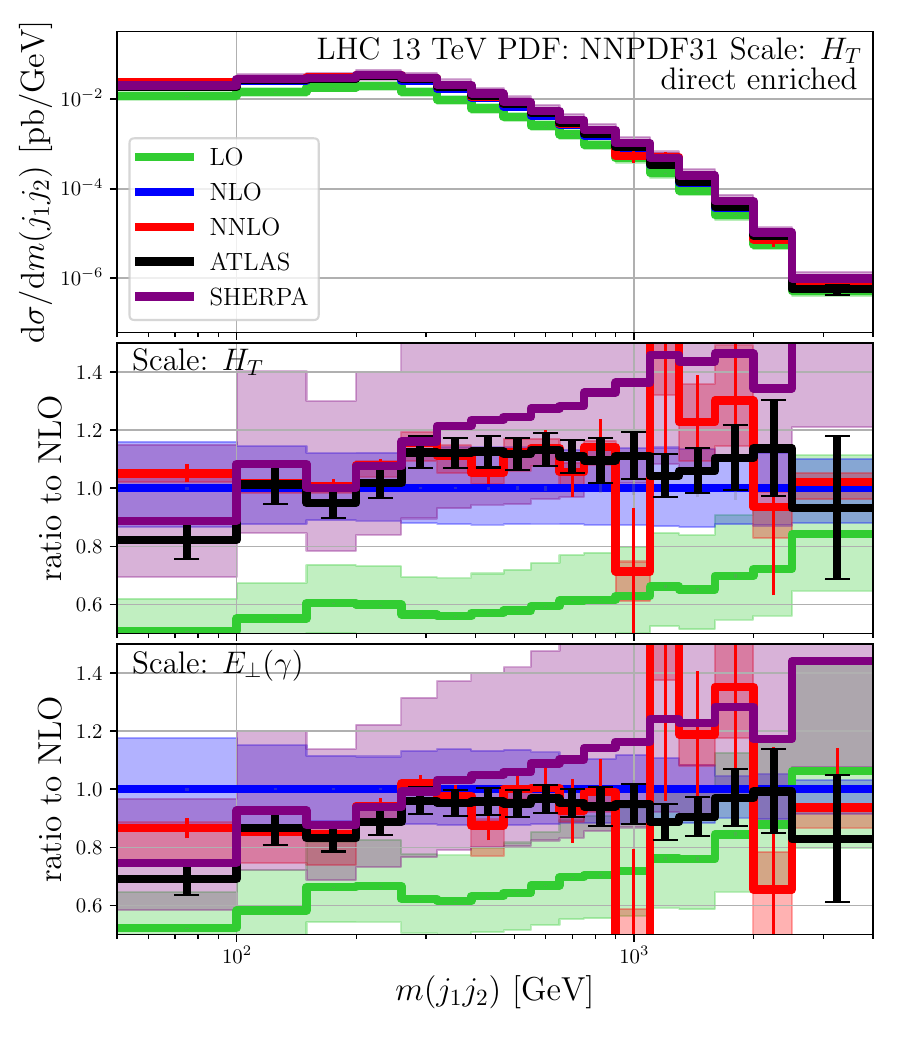}
    \caption{The same as \cref{fig:diffxs_ET} but for the $m(j_1j_2)$ observable.}
    \label{fig:diffxs_mjj}
\end{figure}

The differential cross section is presented in \cref{fig:diffxs_mjj}. Unsurprisingly, the pattern of perturbative corrections is similar to the previous two cases. The NNLO QCD corrections are small and lead to a significant reduction of the scale dependence. The NNLO QCD predictions agree with the data across the spectrum. Only for small invariant masses do we observe a clear deviation. This deviation might be related to the different definition of the photon isolation between theory and experiment. While it seems that \textsc{Sherpa} describes the data in the first bin better than fixed-order QCD, we point out the large fluctuations of the results of the \textsc{Sherpa} simulation, which makes any precise statement impossible. Furthermore, we expect that resummation effects play no role in the first bin of the distribution. This assessment is supported in particular by the fact that the NLO and NNLO predictions with the $H_T$ scale are very close to each other. For high invariant mass, the \textsc{Sherpa} predictions completely fail to describe the spectrum, which is noteworthy because this observable is of special interest in searches for New Physics, e.g.\ in the search for a heavy resonance decaying into jets and recoiling against the prompt photon. The NNLO QCD predictions correctly describe the falling spectrum with only small corrections w.r.t.\ NLO QCD. The scale-variation uncertainty is smaller than the experimental uncertainties. The good description of the data, in particular in comparison to the \textsc{Sherpa} spectrum, motivates the usage of the predicted spectrum instead of relying on data-driven background estimates. Of course, this would require a substantial investment of computing resources in order to improve the quality of the Monte-Carlo integration for invariant masses in the TeV range. In this region, we observe large cancellations between different contributions to the complete NNLO result, which impact the quality of the histograms. The cancellations are enhanced for the $E_\perp(\gamma)$ scale. The results in the {\it direct-enriched} phase space suffer further from reduced statistics due to the additional cuts.

\paragraph{Azimuthal separation between the jets and photon $|\Delta\phi^{\gamma-\text{jet}}|$ and $|\Delta\phi^{j_1-j_2}|$.}

\begin{figure}
    \centering
    \includegraphics[width=0.5\textwidth,page=5]{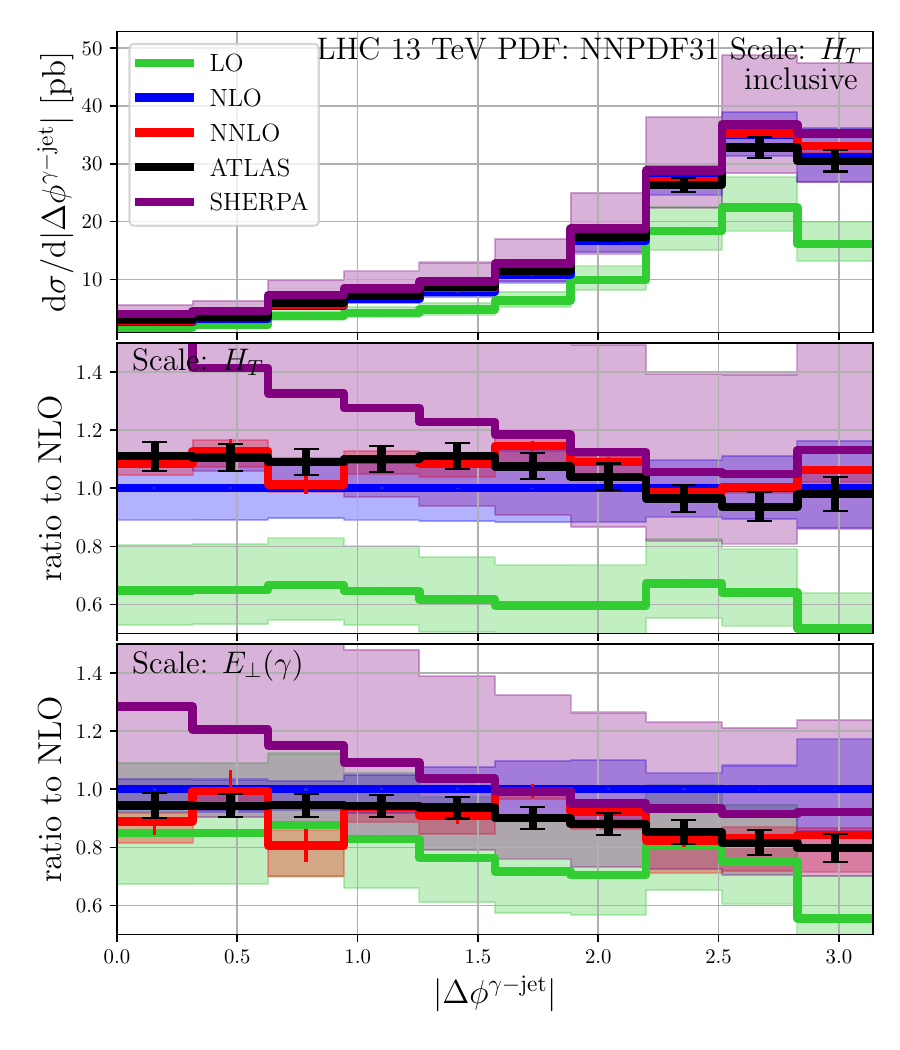}%
    \includegraphics[width=0.5\textwidth,page=7]{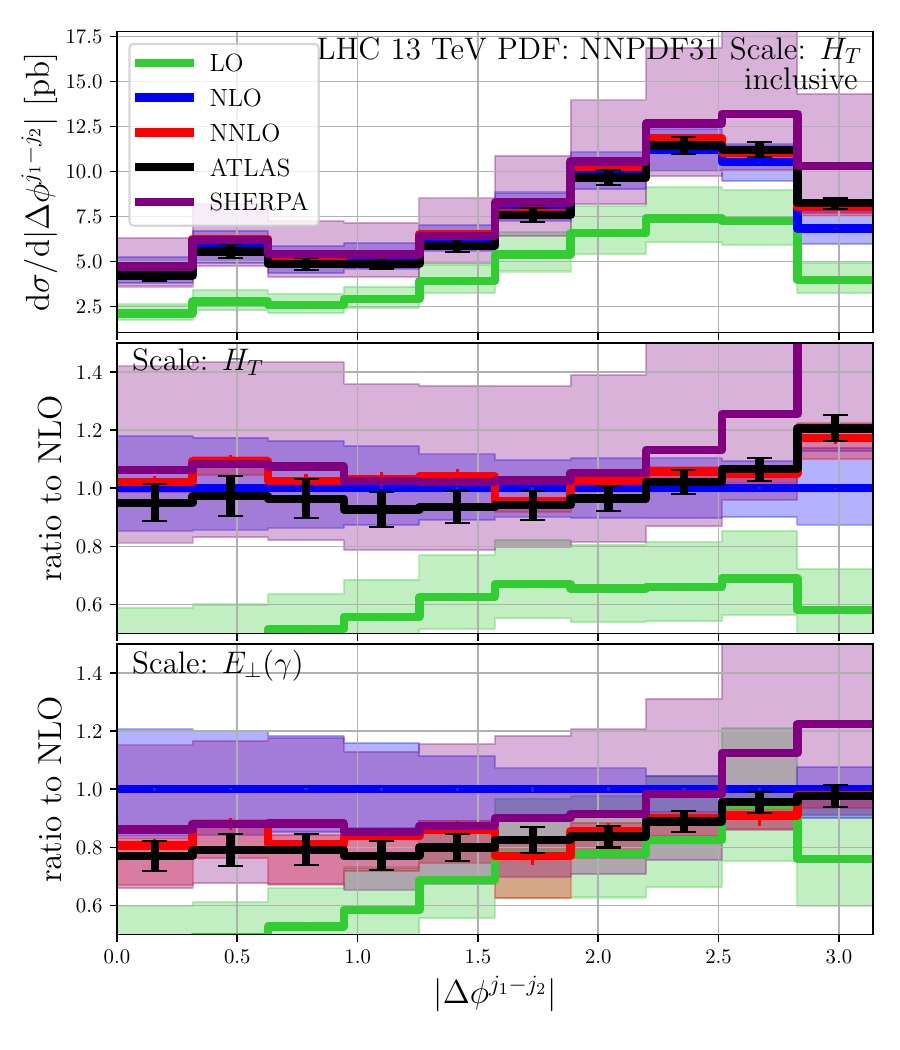}
    \caption{The same as \cref{fig:diffxs_ET} but for the $|\Delta\phi^{\gamma-\text{jet}}|$ and  $|\Delta\phi^{j_1-j_2}|$ observables in the {\it inclusive} phase space only.}
    \label{fig:diffxs_dp}
\end{figure}

These are the last two observables that we discuss in detail. We first focus on the {\it inclusive} phase space, for which the differential cross sections are shown in \cref{fig:diffxs_dp}. For the $H_T$ scale, the NLO QCD predictions agree with the data reasonably well but the NNLO QCD corrections are essential in order to precisely describe the shape of the distributions. The second-order corrections improve the data to theory comparison especially for large azimuthal separations. Interestingly, \textsc{Sherpa} simulations can hardly describe the photon-jet azimuthal distance for small separation, and the jet-jet azimuthal distance for large separation. Both cases correspond to the situation where a photon and a nearby softer jet are likely recoiling against a harder jet, which leads simultaneously to a small $|\Delta\phi^{\gamma-j_2}|$ and large $|\Delta\phi^{j_1-j_2}|$. This is also the region of large $m(j_1j_2)$.

\begin{figure}
    \centering
    \includegraphics[width=0.5\textwidth,page=5]{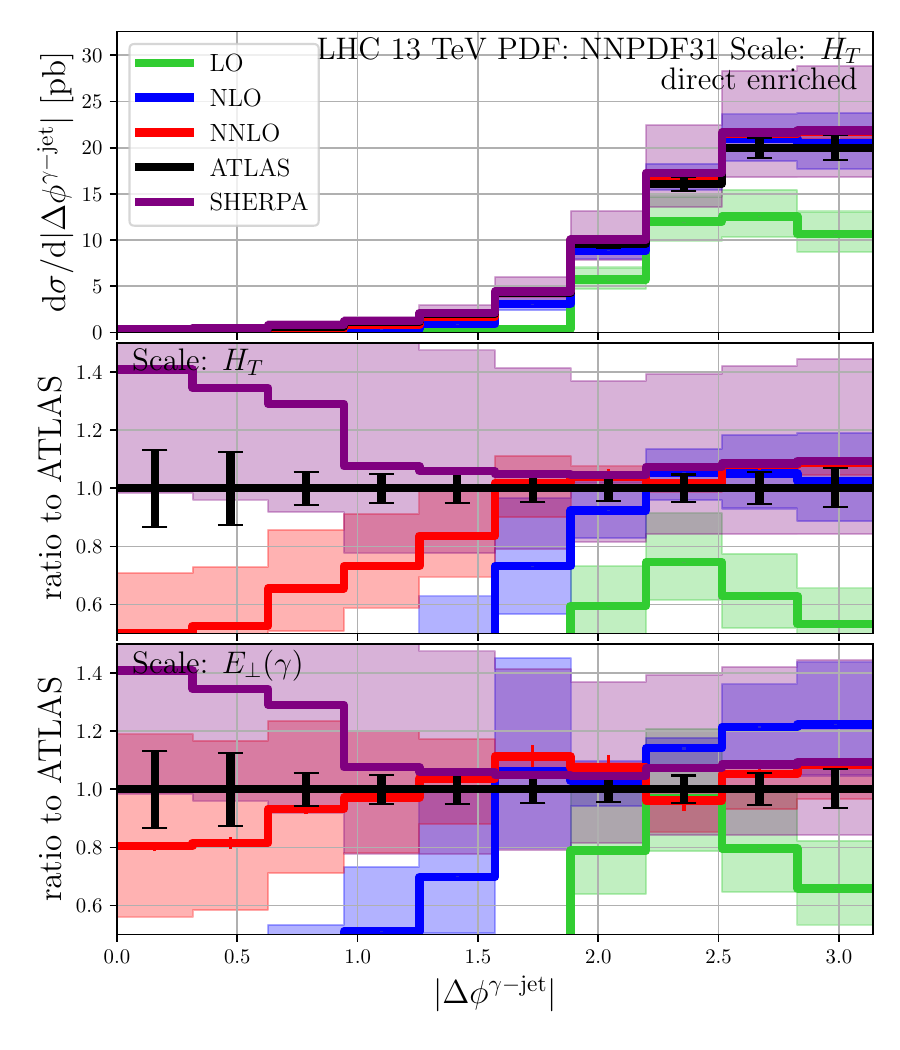}%
    \includegraphics[width=0.5\textwidth,page=7]{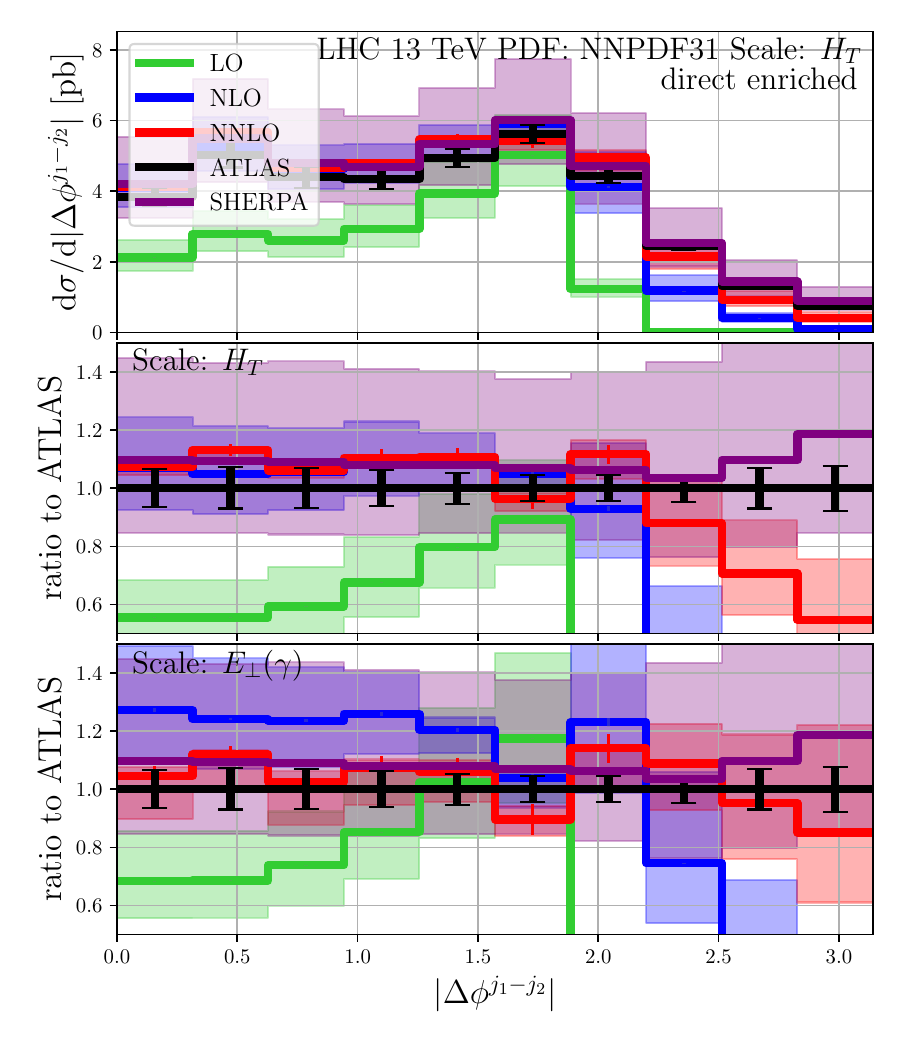}
    \caption{The same as \cref{fig:diffxs_ET} but for the $|\Delta\phi^{\gamma-\text{jet}}|$ and  $|\Delta\phi^{j_1-j_2}|$ observables in the {\it direct-enriched} phase space only. Note that, in this case, we show the ratio plot with respect to data since the perturbative corrections are large. See text for more details.}
    \label{fig:diffxs_dp_prom}
\end{figure}

In the case of the {\it direct-enriched} phase space, the considered observables have some peculiar properties due to the geometric constraints arising from the requirement $E_{\perp}(\gamma) > p_T(j_1)$. The distance between the photon and the jets corresponds then to the same observable which is studied in the context of the azimuthal decorrelation in three jet production \cite{ATLAS:2018sjf}. The additional phase space constraint leads to the fact that $|\Delta\phi^{\gamma-\text{jet}}| > \frac{2 \pi}{3}$ when considering LO kinematics. The presence of a kinematical edge is responsible for large perturbative corrections, as visible in \cref{fig:diffxs_dp_prom}. Indeed this is the only case where the $E_{\perp}(\gamma)$ scale choice performs better than $H_T$, since it enhances the radiation of high-$p_T$ jets filling the phase space better.

\paragraph{Estimate of omitted fragmentation contributions.}

The fragmentation contributions to the differential cross sections presented above may be evaluated by convolving parton-to-photon fragmentation functions with a three-jet production cross section defined with a corresponding resolved parton. The hard-cone isolation criterion, Eq.~\eqref{eq:hardcone}, implies that only hard photons with fragmentation-momentum fraction $z \geq E_\perp(\gamma)/(E_\perp(\gamma) + E_{\perp,{\rm max}}) > 0.93$ contribute. Fragmentation functions are strongly suppressed for large $z$, which is responsible for the dominance of well-separated photon emissions. This effect is further strengthened by the cut requiring the photon to be harder than the hardest jet.

While it is impossible to provide a back-of-the-envelope estimate of the omitted fragmentation contributions, which have been replaced by contributions defined by Frixione's isolation, we may nevertheless invoke the recent Ref.~\cite{Chen:2022gpk}. There, the fragmentation contributions have been included for the photon + jet process with a phase space very similar to ours, since referring to similar measurements by ATLAS. In particular, Figure 9 in Section 5.1 of Ref.~\cite{Chen:2022gpk} shows a comparison between the transverse energy/momentum spectra with fragmentation contributions and with a hybrid photon-isolation criterion. The difference never exceeds 5\%. In view of these findings, we also estimate that the effects omitted in our study should not exceed 5\% in neither the {\it inclusive} nor the {\it direct-enriched} phase space.

The effect is by design larger in the {\it fragmentation-enriched} part of the phase space. Hence, we do not show any comparison of our predictions to data for such configurations. However, for reference, we provide the corresponding plots in the ancillary files attached to the arXiv preprint.

\subsection{Size of the sub-leading-colour terms of the double-virtual contributions}\label{subsec:vvf}

\begin{figure}
    \centering
    \includegraphics[width=0.5\textwidth,page=1]{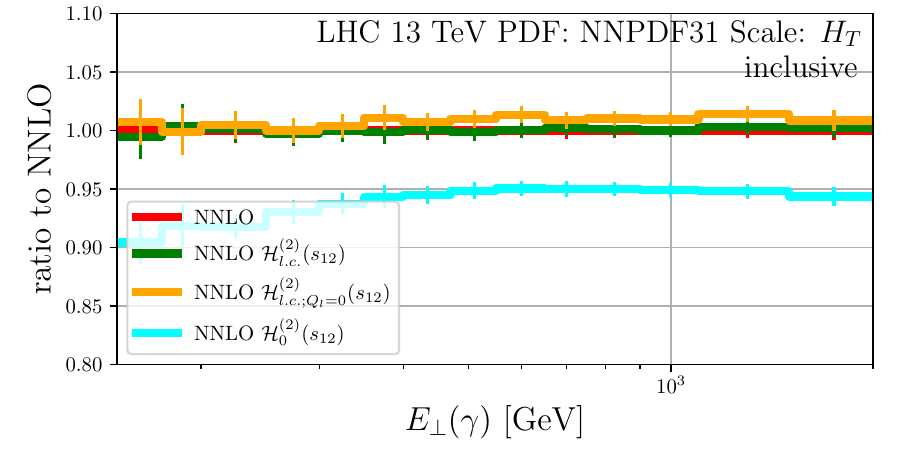}%
    \includegraphics[width=0.5\textwidth,page=2]{plots/nnlo_vva_incl_NNPDF31_HT.pdf}
    \includegraphics[width=0.5\textwidth,page=5]{plots/nnlo_vva_incl_NNPDF31_HT.pdf}%
    \includegraphics[width=0.5\textwidth,page=8]{plots/nnlo_vva_incl_NNPDF31_HT.pdf}
    \caption{A comparison of different approximations to the double virtual corrections. See text for details.}
    \label{fig:vva}
\end{figure}

The differential cross sections presented in the previous Section are the first NNLO QCD results for a two-to-three process with exact double-virtual corrections. Previous calculations for three-photon, di-photon + jet, and three-jet production processes relied on the leading-colour approximation of the two-loop finite remainder, possibly augmented with some sub-leading contributions given by additional planar diagrams. In the case of each of these processes, it was argued that the approximation should not have a noticeable impact on the results due to the expected size of the sub-leading-colour contributions in comparison with the scale uncertainty of the NNLO calculation itself. The argument was based on the small size of the finite remainder in comparison with the complete cross section, as well as on the experience-based expectation that sub-leading-colour contributions should amount to about 10\% of the leading-colour contributions. Here, we have the opportunity to assess the quality of the leading-colour approximation using the exact result. To this end, we consider the double-virtual finite remainder cross section
\begin{equation}
  \sigma^{VVF} = \int \text{d} \Phi \, \mathcal{H}^{(2)}(\mu_R^2)\;,
\end{equation}
where $\mathcal{H}^{(2)}(\mu_R^2)$ is defined in \cref{app:benchmark}. Terms depending on the logarithm of the renormalisation scale, $\ln(\mu_R^2/\mu_{R0}^2)$, occurring in the finite remainder $\mathcal{H}^{(2)}(\mu_R^2)$ can be derived using the formulae presented in Refs.~\cite{Becher:2009cu, Becher:2009qa}, see \cref{app:mu}. They require the evaluation of tree-level and one-loop amplitudes, which can be done at full colour for any process. Hence, we decompose the finite remainder as follows
\begin{align} \begin{aligned}
  \mathcal{H}^{(2)}(\mu_R^2) &= \mathcal{H}^{(2)}(s_{12}) + \sum_{i=1}^4 c_i \ln^i\left(\frac{\mu_R^2}{s_{12}}\right)\\
  &\approx \mathcal{H}^{(2)}_{l.c.}(s_{12}) + \sum_{i=1}^4 c_i \ln^i\left(\frac{\mu_R^2}{s_{12}}\right)\;.
\end{aligned} \end{align}
Note that the choice of the reference scale does influence the size of the sub-leading colour effects in $\mathcal{H}^{(2)}(\mu_{R0}^2)$. Here, just as in our previous calculations for the aforementioned photon and jet processes, we take $\mu_{R0}^2 = s_{12}$. This corresponds to the value used in the implementation of the sector-improved residue subtraction scheme in the \textsc{C++} library \textsc{Stripper}. The choice should not lead to unwanted enhancements of $\mathcal{H}^{(2)}(\mu_{R0}^2)$ as long as the final states are well separated. We compare three different approximations:
\begin{enumerate}
\item[1)] $\mathcal{H}^{(2)}_{l.c.}(s_{12})$: the large $N_c$ limit assuming that $n_f$ and $Q_l$ scale with the same rate. This is more than just the planar approximation since the $Q_l$ terms can have non-planar contributions.
\item[2)] $\mathcal{H}^{(2)}_{l.c.;Q_l=0}(s_{12})$: same as 1) but with $Q_l = 0$.
\item[3)] $\mathcal{H}^{(2)}_{0}(s_{12})$: here we just set $\mathcal{H}^{(2)}(s_{12}) = 0$.
\end{enumerate}

A comparison of the differential cross sections (all contributions, not only the double virtual) assuming each of the three approximations is shown in \cref{fig:vva}. We plot the ratio with respect to the full-colour result labelled ``NNLO''. We observe that there is no noticeable difference between the exact and the approximated results in the case of $\mathcal{H}^{(2)}_{l.c.}(s_{12})$. The $\mathcal{H}^{(2)}_{l.c.;Q_l=0}(s_{12})$ approximation has a small impact reaching at most $1\%$ at the cross section level.

Finally, we can use the $\mathcal{H}^{(2)}_{0}(s_{12})$ approximation to estimate the overall contribution of the two-loop hard functions in the NNLO QCD result. This approximation is clearly distinguishable from the full result and indicates that the double-virtual contribution is about $10\%$ across the phase space. In some cases, the double-virtual contribution may be even larger, as demonstrated by the invariant-mass distribution for large $m(j_1j_2)$ values.

As a side note, we point out that the loop-induced corrections from the $gg \to gg\gamma$ one-loop squared matrix elements never exceed $1$ permille relative to NLO QCD.

\section{Conclusions}
\label{sec:conclusions}

We have derived the exact two-loop amplitudes in the four-quark and two-quark-two-gluon channels necessary for the complete description of single photon hadro-production in conjunction with two final-state jets at NNLO in QCD. 
We employed finite-field arithmetic throughout the computation, and reconstructed the analytic expression of the amplitudes using an optimised  algorithm informed by the expected singularity structure, and boosted by a partial fraction decomposition. We optimised the IBP systems with the help of syzygy equations.
%
%
The results are explicit for a minimal set of helicity configurations and partial amplitudes expressed in terms of pentagon functions suitable for efficient numerical evaluation. The complete hard functions required for the squared matrix elements are obtained by a combination of parity conjugation and permutation of the external momenta.
%
%
Each of these operations has been described in detail. We have also paid particular attention to numerical stability, which is achieved with the help of higher-precision floating-point arithmetic in our implementation in \textsc{C++}.
The two-loop amplitudes obtained here are made available in a public repository~\cite{ppajjampsite}.

Equipped with the two-loop hard functions, we have performed a phenomenological analysis of the photon + di-jet production process in the setup corresponding to recent ATLAS measurements at the LHC. We have demonstrated that the NNLO results improve the description of the data in the case of the photon transverse energy, the jet transverse momentum, and the di-jet invariant mass. Nevertheless, there is a noticeable difference between theory and experiment in the case of the photon transverse energy starting around $1 \; \text{TeV}$. This difference increases with energy and can be attributed to electroweak effects not accounted for in our study. This explanation is consistent with the negative sign of electroweak Sudakov corrections typical of high-transverse-momentum processes. In the case of the azimuthal angle between the photon and the leading jet, and the azimuthal angle between the two leading jets, the description is satisfactory in the inclusive case, while perturbation theory fails as expected for exclusive cuts designed to remove fragmentation effects. Finally, we notice that the theoretical predictions obtained from parton-shower-matched and multi-jet-merged simulations generated with \textsc{Sherpa} have larger scale uncertainties and provide a poorer data description than NLO let alone NNLO QCD predictions. One possible explanation of this unsatisfactory situation could be the reliance on a multi-jet-merged sample of leading-order accuracy. We do not exclude the possibility of improving the description by a better setup of the \textsc{Sherpa} simulation. Should this be achievable, it would demonstrate the pitfalls of using the very complicated machinery of Monte-Carlo event generators.

Since our phenomenological study is designed to showcase the two-loop two-to-three virtual corrections, we have not evaluated the fragmentation contributions. These contributions may be omitted thanks to the special set of cuts called here the {\it direct-enriched} phase space. While we estimate the systematic uncertainty implied by the omission as lower than 5\% and decreasing with energy, we intend to include fragmentation in a future study. Finally, we stress that infrared safety of the observables is guaranteed by the use of the well-known Fixione isolation condition in conjunction with the hard-cone isolation used in the experimental setup.

Our study not only completes the set of predictions for two-to-three processes with photons and jets, but is also the first not to rely on the leading-colour approximation of double-virtual contributions. 

\begin{acknowledgments}
The authors would like to thank Alexander Mitov for fruitful discussions. The work of M.C. was supported by the Deutsche Forschungsgemeinschaft under grant 396021762 – TRR 257. 
H.B.H. is supported by STFC consolidated HEP theory grant ST/T000694/1.
R.P. acknowledges support from the Leverhulme Trust and the Isaac Newton Trust. 
This project received funding from the European Research Council (ERC) under the European Union's research and innovation programmes
\textit{New level of theoretical precision for LHC Run 2 and beyond} (grant agreement No.~683211),
\textit{High precision multi-jet dynamics at the LHC} (grant agreement No.~772099), and 
\textit{High-precision multi-leg Higgs and top physics with finite fields} (grant agreement No.~101040760).
This work was performed using the Cambridge Service for Data Driven Discovery (CSD3), part of which is operated by the University of Cambridge Research Computing on behalf of the STFC DiRAC HPC Facility (www.dirac.ac.uk). The DiRAC component of CSD3 was funded by BEIS capital funding via STFC capital grants ST/P002307/1 and ST/R002452/1 and STFC operations grant ST/R00689X/1. DiRAC is part of the National e-Infrastructure. Simulations were performed with computing resources granted by RWTH Aachen University under project p0020025.

\end{acknowledgments}

\appendix

\section{Numerical benchmark for the hard functions}\label{app:benchmark}

In this Appendix we provide numerical values of the hard functions for the $\qqgga$ and $\qqQQa$ partonic subprocesses at a benchmark phase-space point. We define the tree-level, one-loop, and two-loop hard functions as follows
\begin{subequations}
\begin{align}
\cH^{(0)} & = \colhelsum \big|\cM^{(0)}\big|^2\,, \\
\cH^{(1)} & = 2 \, \mathrm{Re} \, \colhelsum \cM^{(0)*} \cR^{(1)}\,, \\
\cH^{(2)} & = 2 \, \mathrm{Re} \,  \colhelsum \cM^{(0)*} \cR^{(2)}
             + \colhelsum \cR^{(1)*} \cR^{(1)}\,.
\end{align}
\label{eq:hardfunctions}%
\end{subequations}
The factors required for averaging over initial state helicities and colours, as well as the relevant symmetry factors in the case of identical final state partons are included in the normalisation. We define the following short-hand notation
\begin{equation}
f_1 f_2 \to f_3 f_4 \gamma  \qquad \Longrightarrow \qquad f_1(-p_1) + f_2(-p_2) \to f_3(p_3) + f_4(p_4) + \gamma(p_5) \,, 
\label{eq:scatteringdef}
\end{equation}
to represent a scattering process involving two incoming partons ($f_1$, $f_2$), and two outgoing partons ($f_3$, $f_4$) in association with a photon.

In \cref{tab:num-hardfunction}, we show the values of the tree-level, one-loop, and two-loop hard functions 
for a selection of partonic scattering channels that contribute to $pp \to \gamma j j$ 
at the following phase-space point, 
\begin{align}
p_1 & = (-500,  0,  0,  -500)\,, \nn
p_2 & = (-500,  0,  0,   500)\,, \nn
p_3 & = (458.57878788544,  169.45322030968,  379.65366207819, -193.50247465025)\,, \label{eq:PSpoint}\\
p_4 & = (364.06662073682, -18.329869293192, -347.70430131937,  106.34960775871)\,, \nn
p_5 & = (177.35459137774, -151.12335101649, -31.949360758832,  87.152866891544)\,, \nonumber
\end{align}
and using the following parameters,
\begin{equation}
\mu_R  = 173.2\,, \qquad\qquad
\alpha_s  = 0.118\,,\qquad\qquad
e  = 0.30795376724\,.
 \label{eq:numericalparameters} 
\end{equation}
 The momenta and the renormalisation scale $\mu_R$ are given in units of GeV.

\renewcommand{\arraystretch}{1.5}
\begin{table}
    \centering
    \begin{tabular}{|c|ccc|}
    \hline
    $\qqgga$ & $\cH^{(0)} \, \left[\mathrm{GeV}^{-2}\right]$ & $\cH^{(1)}/\cH^{(0)}$ & $\cH^{(2)}/\cH^{(0)}$  \\ 
    \hline
    $u\bar{u}\to gg\gamma$  & $1.42202 \cdot 10^{-5}$ & -0.0868074 & 0.0219214 \\ 
    $gg \to u\bar{u}\gamma$ & $2.69842 \cdot 10^{-6}$ & -0.246582  & 0.0412084 \\
    $ug \to ug\gamma$       & $4.53256 \cdot 10^{-5}$ &  0.0299060 & 0.0169971 \\
    $gu \to ug\gamma$       & $2.70402 \cdot 10^{-5}$ &  0.0155916 & 0.00457076 \\
    \hline
    $\qqQQa$ & $\cH^{(0)} \, \left[\mathrm{GeV}^{-2}\right]$ & $\cH^{(1)}/\cH^{(0)}$ & $\cH^{(2)}/\cH^{(0)}$  \\
    \hline
    $u\bar{u} \to u\bar{u}\gamma$ & $4.29458 \cdot 10^{-5}$ & 0.0756652 & -0.0388519 \\ 
    $u\bar{u} \to d\bar{d}\gamma$ & $4.41265 \cdot 10^{-6}$ & -0.253477 &  0.0493917 \\
    $u\bar{u} \to c\bar{c}\gamma$ & $8.04351 \cdot 10^{-6}$ & -0.290635 &  0.0616392 \\
    $uu \to uu\gamma$             & $1.88278 \cdot 10^{-5}$ &  0.171349 &  0.0140249 \\
    $ud \to ud\gamma$             & $2.00840 \cdot 10^{-5}$ &  0.233819 & -0.0167785 \\
    $uc \to uc\gamma$             & $6.53549 \cdot 10^{-6}$ &  0.228568 & -0.0168878 \\
    \hline
    \end{tabular}
    \caption{Numerical results for the hard functions defined in \cref{eq:hardfunctions} for a selection of partonic reactions contributing to NNLO QCD computation of
    $pp\to\gamma j j$ production at the LHC. We employ the phase-space point given in \cref{eq:PSpoint} and parameters specified in \cref{eq:numericalparameters}.
    The notation used for the partonic scattering process is defined in \cref{eq:scatteringdef}.
    }
    \label{tab:num-hardfunction}
\end{table}

Numerical results for the hard functions for all the scattering channels contributing to $pp\to\gamma j j$, 
as well as all partial helicity finite remainders for all 
contributing helicity configurations can be accessed by running the scripts \verb=SquaredFiniteRemainder_2q2gA.wl= and 
\verb=SquaredFiniteRemainder_2q2QA.wl= available in the public repository~\cite{ppajjampsite}.
These two files also provide all the necessary rules to perform parity transformation and permutations to obtain the full set of helicity configurations, partial amplitudes and scattering channels.

\section{\texorpdfstring{$\mu_R$}{Scale} dependence of the finite remainder}
\label{app:mu}
In the computation of the finite remainders $\mathcal{R}^{(L)}$ we set $\mu_R=1$. We recover the dependence on $\mu_R$ a posteriori, by adding suitable $\mu_R$-restoring terms $\delta\mathcal{R}^{(L)}$ to the finite remainders,
\begin{align}
\mathcal{R}^{(L)}(\vec{s} , \mu_R^2) = \mathcal{R}^{(L)}(\vec{s} , \mu_R^2=1) + \delta\mathcal{R}^{(L)}(\vec{s}, \mu_R^2) \,.
\end{align}
$\vec{s}$ is the set of kinematic invariants defined in \cref{eq:sijs}. The $\mu_R$-restoring terms depend only on the $\beta$ function, the IR operators $\textbf{Z}^{(L)}$, and the lower-loop finite remainders evaluated at $\mu_R^2=1$. Computing the finite remainders with $\mu_R^2=1$ is thus sufficient to recover the full dependence. Explicitly, the one- and two-loop $\mu_R$-restoring terms are given by
\begin{align} \label{eq:mu1L}
& | \delta\mathcal{R}^{(1)}(\vec{s},\mu_R^2) \rangle =  \left[ \frac{1}{2} \log^2(\mu_R^2) \, \textbf{Z}^{(1)}_{-2} +
  \log(\mu_R^2) \left(\beta_0 \, \textbf{I} +\textbf{Z}^{(1)}_{-1}(\vec{s}) \right) \right] | \mathcal{M}^{(0)}(\vec{s}) \rangle \,, \\
\label{eq:mu2L} & | \delta\mathcal{R}^{(2)}(\vec{s},\mu_R^2) \rangle = \left[ \frac{1}{2} \log^2(\mu_R^2)  \, \textbf{Z}^{(1)}_{-2} + \log(\mu_R^2) \left(2 \beta_0 \,  \textbf{I} + \textbf{Z}^{(1)}_{-1}(\vec{s}) \right) \right]  | \mathcal{R}^{(1)}(\vec{s},\mu_R^2=1) \rangle \nonumber \\
& \phantom{ |  \delta\mathcal{R}^{(2)}(\vec{s},\mu_R^2) \rangle = } + 
\biggl\{ \log^4(\mu_R^2) \left[-\frac{5}{24}  \left(\textbf{Z}^{(1)}_{-2}\right)^2+\frac{2}{3} \, \textbf{Z}^{(2)}_{-4} \right] \nonumber \\ 
& \phantom{ | \delta\mathcal{R}^{(2)}(\vec{s},\mu_R^2) \rangle = + \biggl\{} +
  \frac{1}{6} \log^3(\mu_R^2) \left(10 \beta_0 \, \textbf{Z}^{(1)}_{-2}-5 \, \textbf{Z}^{(1)}_{-2} \cdot \textbf{Z}^{(1)}_{-1}(\vec{s}) +8 \, \textbf{Z}^{(2)}_{-3}(\vec{s}) \right) \nonumber \\
& \phantom{ |  \delta\mathcal{R}^{(2)}(\vec{s},\mu_R^2) \rangle = + \biggl\{} +
   \log^2(\mu_R^2) \left[\beta_0^2 \, \textbf{I}+\frac{5}{2} \beta_0 \, \textbf{Z}^{(1)}_{-1}(\vec{s})-\frac{1}{2} \left(\textbf{Z}^{(1)}_{-1}(\vec{s})\right)^2+2 \, \textbf{Z}^{(2)}_{-2}(\vec{s}) \right]+ \nonumber \\
 & \phantom{ | \delta\mathcal{R}^{(2)}(\vec{s},\mu_R^2) \rangle = + \biggl\{} + \log(\mu_R^2) \left( \beta_1 \, \textbf{I} +2 \, \textbf{Z}^{(2)}_{-1}(\vec{s}) \right) \biggr\}  | \mathcal{M}^{(0)}(\vec{s}) \rangle \,,
\end{align}
where $\textbf{I}$ is the identity operator in colour space, and $\textbf{Z}^{(L)}_{k}$ is the order-$\eps^k$ term in the Laurent expansion around $\eps=0$ of the IR operators at $\mu_R^2=1$,
\begin{align}
\textbf{Z}^{(L)}(\vec{s},\mu_R^2=1) = \sum_{w=-2 L}^{-1} \textbf{Z}^{(L)}_{w}(\vec{s}) \, \eps^w \,.
\end{align}
Note that ---~in the scheme we adopted~--- the IR operators do not contain finite terms, i.e.\ they have no terms of order $\eps^0$ or higher. We further recall that the IR operators $\textbf{Z}^{(L)}$ are process-dependent, and that the leading poles ---~$\textbf{Z}^{(L)}_{-2 L}$~--- are constant.
In \cref{eq:mu1L,eq:mu2L} we spelled out the dependence on the kinematic invariants $\vec{s}$, and all logarithms of $\mu_R^2$ are explicit. 

We verify that the finite remainders have the correct scale dependence by evaluating them at two phase-space points related by a rescaling by some factor $a>0$, and checking that they satisfy the relation 
\begin{align}
\frac{\mathcal{R}^{(L)}(a \, \vec{s}, 1 ) + \delta\mathcal{R}^{(L)}(a \, \vec{s}, a )}{
  \mathcal{R}^{(L)}(\vec{s},1) } = \frac{\mathcal{M}^{(0)}(a\, \vec{s}) }{\mathcal{M}^{(0)}(\vec{s})} \,.
\end{align}
This follows from the fact that amplitudes and finite remainders are homogeneous functions of $\vec{s}$ and $\mu_R^2$.

\section{Parity transformation and external-momentum permutation for the amplitudes} \label{app:parity}

We derived analytic expressions for the independent helicity configurations and the minimal set of partial amplitudes. 
In order to obtain the expressions for the complete set of helicity configurations and partial amplitudes required to evaluate the squared matrix elements in the various scattering channels, we apply a parity transformation and/or a permutation of the external momenta to the minimal set of partial amplitudes.
In this Appendix we spell out the algorithms required to perform these operations. We first define the following short-hand notation for the partial amplitudes:
\begin{equation}
\cA^{\vec{h}}(g)  = \cA ^{(L)}\left(1^{h_1},2^{h_2},3^{h_3},4^{h_4},5^{h_5}_{\gamma}\right)\,,
\end{equation}
where $g = (12345)$ is the default external-momentum ordering specified in \cref{eq:2q2gAlabel,eq:2q2QAlabel}, and $\vec{h} = (h_1,h_2,h_3,h_4,h_5)$.
Our analytic expressions for the loop partial amplitudes have the form
\begin{equation} \label{eq:Aform}
\cA^{\vec{h}}(g) = \sum_k c^{\vec{h}}_k (\vec{x}(g)) \, \mathrm{mon}_k(f(g),\mathfrak{c}) \,.
\end{equation}
Here, we extend the notation for the momentum-twistor variables $\vec{x}$ and pentagon functions $f$ to show explicitly their dependence on the routing of the external momenta: $\vec{x}(g)$ and $f(g)$.
$\vec{x}(g)$ indicates an evaluation of the right-hand-side of \cref{eq:momtwistor} at a given phase-space point with external-momentum ordering $g$.
Since the spinor-phase information is lost when using momentum-twistor variables, we restore it in our analytic expressions for the partial amplitudes through suitable helicity-dependent spinor-phase factors $\Phi^{\vec{h}}$ (defined in \cref{eq:2q2gAphase,eq:2q2QAphase} for $\qqgga$ and $\qqQQa$, respectively). In other words, we divide the coefficients $c_k^{\vec{h}}$ in \cref{eq:Aform} by the phase factor expressed in terms of momentum-twistor variables $\Phi^{\vec{h}}(\vec{x})$, and multiply them by the same expressed in terms of spinor products $\Phi^{\vec{h}}(\la ij \ra,[ij])$. We thus have
\begin{equation}
\cA^{\vec{h}}(g) =  \Phi^{\vec{h}}(\la ij \ra,[ij];g) \, \sum_k  \frac{c^{\vec{h}}_k(\vec{x}(g))}{\Phi^{\vec{h}}(\vec{x}(g))} \, \mathrm{mon}_k(f(g),\mathfrak{c}) \,,
\end{equation}
where we spell out the dependence on the external-momentum ordering $g$. The benefit of this approach is that the momentum-twistor dependent coefficients 
$c^{\vec{h}}_k(\vec{x}(g))/\Phi^{\vec{h}}(\vec{x}(g))$ are phase-free quantities, and the phase of the partial amplitude is explicitly restored. This allows us to perform the following two operations.

\begin{itemize}

\item \textbf{Parity transformation} \\
The action of parity $\mathbf{P}$ flips the helicities and the sign of all pseudo-scalar quantities. In particular, it flips the sign of $\tr_5$ and thus exchanges $\tr_+$ with $\tr_-$. Thus, there is
\begin{align}
\cA^{-\vec{h}}(g) & = \mathbf{P} \, \cA^{\vec{h}}(g) \nn
& = \Phi^{\vec{h}}(\la ij \ra,[ij];g)\big|_{\la ij \ra \leftrightarrow [ji]} \, \sum_k \frac{c^{\vec{h}}_k(\vec{x}(g))}{\Phi^{\vec{h}}(\vec{x}(g))}\bigg|_{\trp \leftrightarrow \trm} 
\, \mathrm{mon}_k\left(\mathbf{P}f(g),\mathbf{P}\mathfrak{c} \right) \,.
\end{align}
We recall from \Cref{sec:pfuncs_permutations} that parity acts on the pentagon functions $f$ and the associated transcendental constants $\mathfrak{c}$ by flipping the sign of all parity-odd functions/constants.

Complex conjugation needed in order to construct the squared matrix elements (\cref{eq:qqgga_squared,eq:qqqqa_squared}) requires that ---~on top of the parity transformation~--- we also complex-conjugate the values of the pentagon functions and constants,
\begin{align}
\left(\cA^{\vec{h}}(g) \right)^{*}
& = \Phi^{\vec{h}}(\la ij \ra,[ij];g)\big|_{\la ij \ra \leftrightarrow [ji]} \, \sum_k \frac{c^{\vec{h}}_k(\vec{x}(g))}{\Phi^{\vec{h}}(\vec{x}(g))}\bigg|_{\trp \leftrightarrow \trm} 
\, \mathrm{mon}_k\left(\mathbf{P}(f(g))^*,\mathbf{P}(\mathfrak{c})^*\right) \,.
\end{align}

\item \textbf{Permutation of the external momenta} \\
Let $g^{\prime}$ be an ordering of the external momenta different from $g=(12345)$. We denote by $\sigma$ the permutation of the external momenta which takes from $g$ to $g^{\prime}$, and by $\vec{h}^{\prime}$ the permuted helicities,
\begin{equation}
g^{\prime} = \sigma \circ g\,, \qquad \vec{h}^{\prime} = \sigma \circ \vec{h}\,.
\label{eq:permutation}
\end{equation}
The partial amplitude for the ordering $g^{\prime}$ is given by
\begin{align}
\cA^{\vec{h^\prime}}(g^\prime) & =  \Phi^{\vec{h}}(\la ij \ra,[ij];g^\prime) \, \sum_k \frac{c^{\vec{h}}_k(\vec{x}(g^\prime))}{\Phi^{\vec{h}}(\vec{x}(g^\prime))}  
\, \mathrm{mon}_k\left( \sigma \circ f(g), \sigma \circ \mathfrak{c} \right) \,.
\end{align}
We emphasise that some of the constants $\mathfrak{c}$ have odd parity, and change sign if the permutation $\sigma$ has odd signature. A more thorough discussion of the permutation of pentagon functions and transcendental constants can be found in \cref{sec:pfuncs_permutations}.

\end{itemize}

\bibliography{main}
\bibliographystyle{JHEP}

\end{document}